\documentclass[a4paper,12pt]{article}
\usepackage{epsfig}
\usepackage[dvips,usenames]{color}
\usepackage{graphicx}

\newlength{\dinwidth}
\newlength{\dinmargin}
\newcommand{\spur}[1]{\not\! #1 \,}
\def\pslash{\rlap{\hspace{0.02cm}/}{p}}
\def\kslash{\rlap{\hspace{0.02cm}/}{k}}
\begin{document}

\title{\bf Revisiting $ B\to  \pi\pi,\pi K$ Decays in QCD Factorization Approach}
\bigskip

\author{ Xinqiang Li$^{1,2,3}$,~Yadong Yang$^{1}$~\footnote{ Corresponding author.
E-mail address: yangyd@henannu.edu.cn}
\\
{ $^1$\small Department of Physics, Henan Normal University,
Xinxiang, Henan 453007,  P.R. China }
\\
{ $^2$\small Institute of Theoretical Physics, Chinese Academy of
Sciences, Beijing, 100080, P.R. China}
\\
{ $^3$\small Graduate School of the Chinese Academy of Science,
Beijing, 100039, P.~R. China}}

\maketitle
\begin{picture}(0,0)
\put(305,290){\sf hep-ph/0508079}
\end{picture}
\bigskip\bigskip
\maketitle \vspace{-1.5cm}
\begin{abstract}
Motivated by the recent experimental data, we have revisited the
$B\to \pi K,\pi \pi$ decays in the framework of QCD factorization,
with inclusion of the important strong penguin corrections of
order $\alpha_s^2$ induced by $b\to D g^\ast g^\ast$~($D=d$ or $s$
and $g^\ast$ denotes an off-shell gluon) transitions. We find that
these higher order strong penguin contributions can provide $\sim
30\%$ enhancement to the penguin-dominated $B\to \pi K$ decay
rates, and such an enhancement can improve the consistency between
the theoretical predictions and the experimental data
significantly, while for the tree-dominated $B\to \pi\pi$ decays,
these higher order contributions play only a minor role. When
these strong penguin contributions are summed, only a small strong
phase remains and  the direct $CP$ asymmetries get small
corrections. We also find patterns of the ratios between the
$CP$-averaged branching fractions remain nearly unaffected even
after including these higher order corrections and the $\pi K$
puzzle still persists.  Our results may indicate that to resolve
the puzzle one would have to resort to new physics contributions
in the electroweak penguin sector as found by Buras {\it et al}.

\end{abstract}


\newpage
\section{Introduction}

The study of exclusive hadronic $B$-meson decays can provide not
only an interesting avenue to understand the $CP$ violation and
flavor mixing of the quark sector in the Standard Model~(SM), but
also powerful means to probe different new physics scenarios
beyond the SM. With the operation of $B$-factory experiments,
large amount of experimental data on hadronic $B$-meson decays are
being collected and measurements of previously known observables
are becoming more and more precise. Thus, studies of the hadronic
$B$-meson decays have entered a precision era.

With respect to the theoretical aspect, several novel methods have
also been proposed to study exclusive hadronic $B$ decays, such as
the ``naive" factorization~(NF)~\cite{BSW}, the perturbative QCD
method~(pQCD)~\cite{pqcd}, the QCD
factorization~(QCDF)~\cite{bbns1,Neubert}, the soft collinear
effective theory~(SCET)~\cite{scet}, and so on. For quite a long
time, the decay amplitudes for exclusive two-body hadronic $B$
decays were estimated in the NF approach, and in many cases, this
approach could provide the correct order of the magnitude of the
branching fractions. However, it cannot predict the direct $CP$
asymmetries properly due to the assumption of no strong
rescattering in the final states. It is therefore no longer
adequate to account for the new $B$-factory data. The other
methods mentioned above are proposed to supersede this
conventional approach. Since we shall use QCDF approach in this
paper, we would only focus on this approach below.

The essence of the QCDF approach can be summarized as follows:
since the $b$ quark mass is much larger than the strong
interaction scale $\Lambda_{QCD}$, in the heavy quark limit
$m_b\gg\Lambda_{QCD}$, the hadronic matrix elements relevant to
two-body hadronic $B$-meson decays can be represented in the
factorization form~\cite{bbns1}
\begin{eqnarray}\label{fact1}
\langle M_{1}(p_{1}) M_{2}(p_{2}){\mid} Q_i {\mid}B(p)\rangle &=&
\langle M_{1}(p_{1}){\mid} j_1 {\mid} B(p)\rangle \langle
M_{2}(p_2){\mid} j_2 {\mid}0\rangle
\nonumber\\
&& \cdot\Big[1+\sum r_n \alpha_s^n + {\cal O}
(\Lambda_{QCD}/m_b)\Big],
\end{eqnarray}
where $Q_i$ is the local four-quark operator in the effective weak
Hamiltonian, $j_{1,2}$ are bilinear quark currents, and $M_1$ is
the meson that picks up the spectator quark from the $B$ meson,
while $M_2$ is the one that can be factored out from the $(B,M_1)$
system. This scheme has incorporated elements of the NF
approach~(as the leading contribution) and the hard-scattering
approach~(as the sub-leading corrections). It provides a means to
compute the hadronic matrix elements systematically. In
particular, the final-state strong interaction phases, which are
very important for studying $CP$ violation in $B$-meson decays,
are calculable from first principles with this formalism. Its
accuracy is limited only by higher order power corrections to the
heavy-quark limit and the uncertainties of theoretical input
parameters such as quark masses, form factors, and the light-cone
distribution amplitudes. Details about the conceptual foundations
and the arguments of this approach could be found in
Ref.~\cite{bbns1,Neubert}.

Among the two-body hadronic $B$-meson decays, the charmless $B\to
\pi K$ and $B\to \pi\pi$ modes are very interesting, since a
significant interference of tree and penguin amplitudes is
expected, and hence have been studied most extensively.
Experimentally, all the four decay channels for $B\to \pi
K$~($B^{\pm}\to \pi^{\pm}K^0$, $B^{\pm} \to \pi^0 K^{\pm}$,
$B^0\to \pi^{\pm}K^{\mp} $, and $B^0 \to K^0 \pi^0$) and the three
ones for $B\to \pi \pi$~($B^{\pm} \to \pi^{\pm} \pi^0$, $B^0 \to
\pi^+ \pi^-$, and $B^0 \to \pi^0 \pi^0$) have been observed with
the $CP$-averaged branching ratios measured within a few percent
errors by the CLEO~\cite{Cronin-Hennessy,Asner,Bornheim},
BaBar~\cite{babar}, and Belle~\cite{belle} collaborations. The
$CP$ asymmetries in these decay modes have also been measured
recently~\cite{Chen,prl:021601,Aubert,Giorgi,Chao1,Chao2,Abe,prl:181802,
prl:181803}. In particular, measurements of the direct $CP$
asymmetry in $B^0 \to \pi^{\pm}K^{\mp}$ have been recently
achieved at the $5.7\sigma$ level by BaBar~\cite{Aubert,Giorgi}
and Belle~\cite{Chao1,Chao2,Abe,Sakai}. All these experimental
data can therefore provide very useful information for improving
the existing model calculations. On the theoretical side, these
decay modes have also been analyzed in detail within the QCDF
formalism~\cite{bbns2,Muta:2000ti,Du:2000ff,Du:2001hr}. Due to
lack of precise experimental data at that time, no large
discrepancies between the theoretical predictions and the
experimental data were found. However, the current new $B$-factory
data for $B\to \pi K, \pi \pi$ decays indicate some potential
inconsistencies with the predictions based on this scheme. For
example, new experimental data for $B^0 \to \pi^0 K^0, \pi^0
\pi^0$ decay rates are significantly larger than the theoretical
predictions with this scheme. In addition, predictions for the
direct $CP$ asymmetries in these modes are also inconsistent with
the data, even with the opposite sign for some
processes~\cite{bbns2,bbns3}. Moreover, the experimental results
of the following ratios between the $CP$-averaged branching
fractions for $B\to \pi K, \pi \pi$
decays~\cite{FM-charge,BF-neutral1}
\begin{eqnarray}\label{Rdefine1}
R_{+-}&\equiv&2\left[\frac{\mbox{BR}(B^+\to\pi^+\pi^0)
+\mbox{BR}(B^-\to\pi^-\pi^0)}{\mbox{BR}(B_d^0\to\pi^+\pi^-)
+\mbox{BR}(\bar
B_d^0\to\pi^+\pi^-)}\right]\frac{\tau_{B^0_d}}{\tau_{B^+}}=
2.20\pm0.31\,,\\
R_{00}&\equiv&2\left[\frac{\mbox{BR}(B_d^0\to\pi^0\pi^0)+
\mbox{BR}(\bar B_d^0\to\pi^0\pi^0)}{\mbox{BR}(B_d^0\to\pi^+\pi^-)+
\mbox{BR}(\bar B_d^0\to\pi^+\pi^-)}\right]=0.67\pm0.14\,,\\
R&\equiv&\left[\frac{\mbox{BR}(B_d^0\to\pi^- K^+)+ \mbox{BR}(\bar
B_d^0\to\pi^+ K^-)}{\mbox{BR}(B^+\to\pi^+ K^0)+
\mbox{BR}(B^-\to\pi^- \bar K^0)}
\right]\frac{\tau_{B^+}}{\tau_{B^0_d}} =0.82\pm 0.06\,,\\
R_{\rm c}&\equiv&2\left[\frac{\mbox{BR}(B^+\to\pi^0K^+)+
\mbox{BR}(B^-\to\pi^0K^-)}{\mbox{BR}(B^+\to\pi^+ K^0)+
\mbox{BR}(B^-\to\pi^- \bar K^0)}\right]=1.00\pm 0.09\,,\\
R_{\rm n}&\equiv&\frac{1}{2}\left[ \frac{\mbox{BR}(B_d^0\to\pi^-
K^+)+ \mbox{BR}(\bar B_d^0\to\pi^+
K^-)}{\mbox{BR}(B_d^0\to\pi^0K^0)+ \mbox{BR}(\bar
B_d^0\to\pi^0\bar K^0)}\right]=0.79\pm 0.08\,,\label{Rdefine5}
\end{eqnarray}
with numerical results compiled by the Heavy Flavor Averaging
Group~(HFAG)~\cite{HFAG}, have shown very puzzling patterns
 \cite{BF,Buras:2004th}. Within the SM, predictions based on the
QCDF approach give $R_{c}\approx R_{n}$, while the value for $R$
is quite consistent with the experimental data~\cite{bbns3}. The
central values for $R_{+-}$ and $R_{00}$ calculated with the QCD
factorization~\cite{bbns3} give $R_{+-}=1.24$ and $R_{00}=0.07$ as
emphasized by Buras $et~ al.$ \cite{Buras:2004th}, which are also
inconsistent with the current experimental data. Though none of
these exciting results is conclusive at the moment due to large
uncertainties both theoretically and experimentally, it is
important and interesting to take them seriously and to find out
possible origins of these discrepancies. Recently, quite a lot of
works have been done to study the implications of these new
experimental
data~\cite{Buras:2004th,Mishima:2004um,Wu:2004xx,Charng:2004ed,He:2004ck,
Baek:2004rp,Carruthers:2004gj,Nandi:2004dx,Morozumi:2004ea,Burrell,Kim,Yeh}.
In this paper, we restrict ourselves to the possibility that these
deviations result from our insufficient understanding of the
hadronic dynamics and investigate the higher order strong penguin
effects induced by $b\to D g^\ast g^\ast$ transitions, where $D=d$
or $s$, depends on the specific decay modes. The off-shell gluons
$g^\ast$ are either emitted from the internal quark loops,
external quark lines, or splitted off the virtual gluon of the
strong penguin.

As shown in literature
\cite{hou1,hou2,simma,greub,bosch,yang:phiXs}, contributions of
the higher order $b\to s g g$ process to the inclusive and
semi-inclusive decay rates of $B$-meson decays could be large
compared to $b\to s g$ process. For example, in \cite{greub},
Greub and Liniger have found that the next-to-leading logarithmic
result of ${\cal B}^{NLL}(b\to sg)=(5.0\pm1.0)\times 10^{-5}$ is
more than a factor of two larger than the leading logarithmic one
${\cal B}^{LL}(b\to sg)=(2.2\pm 0.8)\times 10^{-5}$. In addition,
in \cite{yang:phiXs}, we have found the higher order strong
penguin could give  large corrections to $B\to\phi X_s$. We also
note that the large higher order chromo-magnetic penguin
contributions have also been found by Mishima and
Sanda\cite{mishima} in PQCD factorization framework. Since the
$B\to\pi K$ decays are dominated by strong penguin contributions,
it is interesting to investigate these higher order $b \to s
g^\ast g^\ast$ strong penguin effects on these penguin-dominated
processes. However, for self-consistent, we will also investigate
these effects on the tree-dominated $B\to \pi\pi$ decays. After
direct calculations, we find that these higher order strong
penguin contributions can provide $\sim 30\%$ enhancement to the
penguin-dominated $B\to \pi K$ decay rates, and such an
enhancement can improve the consistency between the theoretical
predictions and the experimental data effectively. For
tree-dominated $B\to \pi\pi$ decays, however, their effects are
quite small. Since the $b\to D g^\ast g^\ast$ strong penguin
contributions contain only a relatively small strong phase, their
effects on the direct $CP$ asymmetries are also small. In
addition, the patterns of the quantities $R$, $R_c$, $R_n$,
$R_{+-}$, and $R_{00}$ defined above remain unaffected even with
these new contributions included.

This paper is organized as follows: In Sec.2, using the QCDF
approach, we first calculate the $B\to \pi K,\pi \pi$ decay
amplitudes at the next-to-leading order in $\alpha_s$, and then
take into account the $b \to D g^\ast g^\ast$ strong penguin
contributions to the decay amplitudes. In Sec.3, after presenting
the theoretical input parameters relevant to our analysis, we give
our numerical results for $B\to \pi K$ and $B\to \pi\pi$ decays.
Some discussions on these higher order corrections and the
$\gamma$ dependence of the relevant quantities are also presented.
Finally, we conclude with a summary in Sec.4. In Appendix A, we
present the correction functions at next-to-leading order in
$\alpha_s$. Explicit form for the quark loop functions are given
in Appendix B.

\section{Decay amplitudes for $B\to \pi K, \pi\pi$ decays in QCDF approach }

\subsection{The effective weak Hamiltonian for hadronic $B$ decays}

In phenomenological treatment of the hadronic $B$-meson decays,
the starting point is the effective weak Hamiltonian at the low
energy~\cite{Buchalla:1996vs,Buras:1998ra}, which is obtained by
integrating out the heavy degree of freedom (e.g. the top quark,
$W^{\pm}$ and $Z$ bosons in the SM) from the Lagrangian of the
full theory. After using the unitarity relation
$-\lambda_t=\lambda_u+\lambda_c$, it can be written as
\begin{equation}\label{Heff}
   {\cal H}_{\rm eff} = \frac{G_F}{\sqrt2} \sum_{p=u,c} \!
   \lambda_p^{(\prime)} \bigg( C_1\,Q_1^p + C_2\,Q_2^p
   + \!\sum_{i=3,\dots, 10}\! C_i\,Q_i + C_{7\gamma}\,Q_{7\gamma}
   + C_{8g}\,Q_{8g} \bigg) + \mbox{h.c.} \,,
\end{equation}
where $\lambda_p=V_{pb}\,V_{ps}^*$ (for $b\to s$ transition) and
$\lambda_p^\prime=V_{pb}\,V_{pd}^*$ (for $b\to d$ transition) are
products of the CKM matrix elements. The effective operators
$Q_{i}$ govern a given decay process and their explicit form can
be read as follows.
\newline
\noindent - Current-current operators:
\begin{eqnarray}
   Q_1^p &=& (\bar p b)_{V-A} (\bar D p)_{V-A} \,,
    \hspace{2.5cm}
    Q^p_2 = (\bar p_i b_j)_{V-A} (\bar D_j p_i)_{V-A} \,,
\end{eqnarray}
\noindent - QCD-penguin operators:
\begin{eqnarray}
   Q_3 &=& (\bar D b)_{V-A} \sum{}_{\!q}\,(\bar q q)_{V-A} \,,
    \hspace{1.7cm}
    Q_4 = (\bar D_i b_j)_{V-A} \sum{}_{\!q}\,(\bar q_j q_i)_{V-A} \,,
    \nonumber\\
   Q_5 &=& (\bar D b)_{V-A} \sum{}_{\!q}\,(\bar q q)_{V+A} \,,
    \hspace{1.7cm}
    Q_6 = (\bar D_i b_j)_{V-A} \sum{}_{\!q}\,(\bar q_j q_i)_{V+A} \,,
\end{eqnarray}
\noindent - Electroweak penguin operators:
\begin{eqnarray}
   Q_7 &=& (\bar D b)_{V-A} \sum{}_{\!q}\,{\textstyle\frac32} e_q
    (\bar q q)_{V+A} \,, \hspace{1.11cm}
    Q_8 = (\bar D_i b_j)_{V-A} \sum{}_{\!q}\,{\textstyle\frac32} e_q
    (\bar q_j q_i)_{V+A} \,, \nonumber \\
   Q_9 &=& (\bar D b)_{V-A} \sum{}_{\!q}\,{\textstyle\frac32} e_q
    (\bar q q)_{V-A} \,, \hspace{0.98cm}
    Q_{10} = (\bar D_i b_j)_{V-A} \sum{}_{\!q}\,{\textstyle\frac32} e_q
    (\bar q_j q_i)_{V-A} \,,
\end{eqnarray}
\noindent - Electro- and chromo-magnetic dipole operators:
\begin{eqnarray}
   Q_{7\gamma} &=& \frac{-e}{8\,\pi^2}\,m_b\,
    \bar D \,\sigma_{\mu\nu}\,(1+\gamma_5)\, F^{\mu\nu} b \,,
    \hspace{0.81cm}
   Q_{8g} = \frac{-g_s}{8\,\pi^2}\,m_b\,
    \bar D \,\sigma_{\mu\nu}\,(1+\gamma_5)\, G^{\mu\nu} b \,,
\end{eqnarray}
where $(\bar q_1 q_2)_{V\pm A}=\bar
q_1\gamma_\mu(1\pm\gamma_5)q_2$, $i,j$ are colour indices, $e_q$
are the electric charges of the quarks in units of $|e|$, and a
summation over $q=u,d,s,c,b$ is implied. For decay modes induced
by the quark level $b\to d$ transition, $D=d$, while for $b\to s$
transition, $D=s$.

The Wilson coefficients $C_{i}(\mu)$ in Eq.~(\ref{Heff}) represent
all the contributions from physics with scale higher than $\mu\sim
{\cal O}(m_{b})$ and have been reliably evaluated up to the
next-to-leading logarithmic order. Numerical results for these
coefficients evaluated at different scales can be found in
\cite{Buchalla:1996vs}.

\subsection{Decay amplitudes at the next-to-leading order in $\alpha_s$}

\begin{figure}[t]
\epsfxsize=10cm \centerline{\epsffile{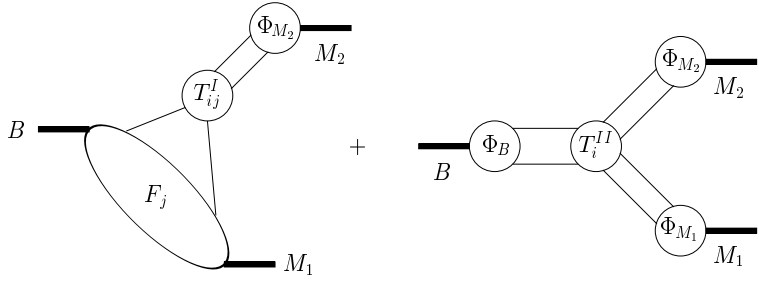}}
\centerline{\parbox{15cm}{\caption{\label{fact-fig} Graphical
representation of the factorization formula. Only one of the two
form-factor terms in (\ref{fact}) is shown for simplicity.}}}
\end{figure}

Using the weak effective Hamiltonian given by Eq~(\ref{Heff}), we
can now write the decay amplitudes for the general two-body
hadronic $B\to M_1 M_2$ decays as
\begin{equation}\label{fac1}
   \langle M_1 M_2|{\cal H}_{\rm eff}|B\rangle
   = \frac{G_F}{\sqrt2} \sum_{p=u,c} \lambda_p\,C_i\,
   \langle M_1 M_2|Q_i^p|B\rangle \,.
\end{equation}
Then, the most essential theoretical problem obstructing the
calculation of the hadronic $B$-meson decay amplitudes resides in
the evaluation of the hadronic matrix elements of the local
operators $\langle M_1 M_2|Q_i^p|B\rangle$. Within the formalism
of the QCDF, this quantity could be simplified greatly in the
heavy-quark limit. To leading power in $\Lambda_{\rm QCD}/m_b$,
but to all orders in perturbation theory, it obeys the following
factorization formula~\cite{bbns2}
\begin{eqnarray}\label{fact}
 \langle M_1 M_2|Q_i^p|B\rangle &=&
 F_j^{B\to M_1}(m_{M_2}^2)\,\times T_{M_2,ij}^{\rm I}*\Phi_{M_2} + F_j^{B\to
 M_2}(m_{M_1}^2)\,\times T_{M_1,ij}^{\rm I}*\Phi_{M_1}
 \nonumber\\
 &&\mbox{}+ T_i^{\rm II}*\Phi_B*\Phi_{M_1}*\Phi_{M_2} \,,
\end{eqnarray}
where $\Phi_M$ is the leading-twist light-cone distribution
amplitude of the meson $M$, and the $*$ products indicate an
integration over the light-cone momentum fractions of the
constituent quarks inside the mesons. The quantity $F_j^{B\to M}$
denotes the $B\to M$ transition form factor. This formula is
illustrated by the graphs shown in Fig.~\ref{fact-fig}.

\begin{figure}[t]
\epsfxsize=12cm \centerline{\epsffile{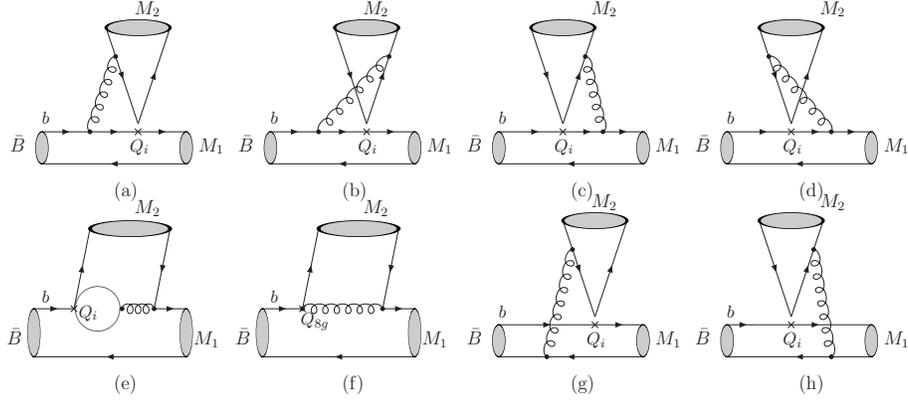}}
\centerline{\parbox{15cm}{\caption{\label{asfig} Order $\alpha_s$
corrections to the hard-scattering kernels $T_{M,ij}^{\rm I}$
(coming from the diagrams (a)-(f)) and $T_i^{\rm II}$ (from the
last two diagrams).}}}
\end{figure}

In Eq.~(\ref{fact}), the hard-scattering kernels $T_{M,ij}^{\rm
I}$ and $T_i^{\rm II}$ are calculable order by order with the
perturbation theory. $T_{M,ij}^{\rm I}$ starts at tree level, and
at higher order in $\alpha_s$ contains the ``non-factorizable''
corrections from the hard gluon exchange and the light-quark
loops~(penguin topologies). The hard ``non-factorizable''
interactions involving the spectator quark are part of the kernel
$T_i^{\rm II}$. At the leading order, $T_{M,ij}^{\rm I}=1$,
$T_i^{\rm II}=0$, and the QCDF formula reproduce the NF results.
Nonperturbative effects are either suppressed by
$\Lambda_{QCD}/m_b$ or parameterized in terms of the meson decay
constants, the transition form factors $F_j^{B\to M}$, and the
light-cone distribution amplitudes $\Phi_B$, $\Phi_M$. The
relevant Feynman diagrams contributing to these kernels at the
next-to-leading in $\alpha_s$ are shown in Fig.~\ref{asfig}.

\begin{figure}[t]
\epsfxsize=13cm \centerline{\epsffile{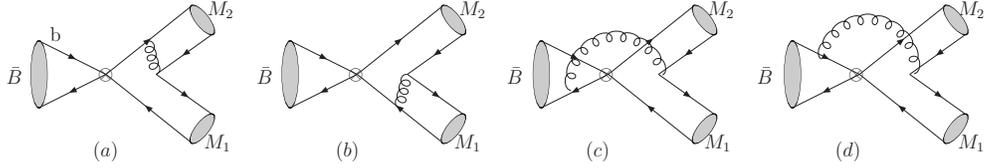}}
\centerline{\parbox{15cm}{\caption{\label{annhfig} The
annihilation diagrams of order $\alpha_s$.}}}
\end{figure}

According to the arguments in \cite{bbns1}, the weak annihilation
contributions to the decay amplitudes are power suppressed
compared to the leading spectator interaction in the heavy quark
limit, and hence do not appear in the factorization
formula~(\ref{fact}). Nevertheless, as emphasized in
\cite{pqcd,CDLU,Kagan}, these contributions may be numerically
important for realistic $B$-meson decays. In particular, the
annihilation contributions with QCD corrections could give
potentially large strong phases, hence large $CP$ violation could
be expected~\cite{pqcd,CDLU}. It is therefore necessary to take
these annihilation contributions into account. At leading order in
$\alpha_s$, the annihilation kernels arise from the four diagrams
shown in Fig.~\ref{annhfig}. They result in a further contribution
to the hard-scattering kernel $T_i^{\rm II}$ in the factorization
formula.

As indicated in the factorization formula~(\ref{fact}), the meson
light-cone distribution amplitudes~(LCDAs) play an important role
in the QCDF formalism. For convenience, we list the relevant
formula as follows (details can be found in ~\cite{Beneke:2000wa})
\newline
\noindent {\bf - LCDAs for $B$ meson.} In the heavy quark limit,
the light-cone projector for the $B$ meson in the momentum space
can be expressed as~\cite{bbns1,Beneke:2000wa,Grozin:1996pq}
 \begin{equation}
 {\cal M}_{\alpha\beta}^B=-\frac{i f_B\, m_B}{4}\,\left[\,(1+\spur{v}
 )\,\gamma_5 \,\left\{\Phi_1^B (\xi) +\spur{n_-} \Phi_2^B
(\xi)\,\right\}\, \right]_{\beta\alpha}\,,
 \label{Bprojector}
 \end{equation}
with the normalization condition
 \begin{equation}
 \int_{0}^{1}\!{d\xi}\,\Phi_1^B (\xi)=1,\,\qquad\qquad
 \int_{0}^{1}\!{d\xi}\,\Phi_2^B (\xi)=0\,,
 \end{equation}
where $\xi$ is the momentum fraction of the spectator quark in the
$B$ meson. For simplicity, we consider only the leading twist
$\Phi_1^B(\xi)$ contribution in this paper. Since almost all the
momentum of the $B$ meson is carried by the heavy $b$ quark, we
expect that $\Phi_1^B(\xi)={\cal O}(m_b/\Lambda_{\rm QCD})$ and
$\xi={\cal O}(\Lambda_{\rm QCD}/m_b)$.
\newline
\noindent {\bf - LCDAs for light mesons.} For the light-cone
projector of light pseudoscalar mesons in momentum space, we use
the form given by~\cite{Geshkenbein:qn}
\begin{equation}\label{projector}
 M_{\alpha\beta}^P = \frac{i f_P}{4} \Bigg\{
   \pslash\,\gamma_5\,\Phi(x) - \mu_P\gamma_5
   \frac{\kslash_2\,\kslash_1}{k_1\cdot k_2}\,\Phi_p(x)
   \Bigg\}_{\alpha\beta},
\end{equation}
where $f_P$ and $p$ are the decay constant and the momentum of the
meson. The parameter
$\mu_P=m_P^2/(\overline{m}_1(\mu)+\overline{m}_2(\mu))$, with
$\overline{m}_{1,2}(\mu)$ being the current quark mass of the
meson constituents, is proportional to the chiral quark
condensate. $\Phi(x)$ is the leading-twist distribution amplitude,
whereas $\Phi_p(x)$ the sub-leading twist (twist-3) one. All of
them are normalized to $1$. The quark and anti-quark momenta of
meson constituents, $k_1$ and $k_2$, are defined respectively by
\begin{equation}\label{momenta2}
   k_1^\mu = x p^\mu + k_\perp^\mu
    + \frac{\vec k_\perp^2}{2x p\cdot\bar p}\,\bar p^\mu \,, \qquad
   k_2^\mu = (1-x)\, p^\mu - k_\perp^\mu
    + \frac{\vec k_\perp^2}{2\,(1-x)\, p\cdot\bar p}\,\bar p^\mu\,,
\end{equation}
where $\bar p$ is a light-like vector whose 3-components point
into the opposite direction of $\vec{p}$. It is understood that
only after the factor $k_1\cdot k_2$ in the denominator of
Eq.~(\ref{projector}) cancelled, can we take the collinear
approximation, i.e., the momentum $k_1$ and $k_2$ can be set to $x
p$ and $(1-x)\, p$, respectively.

From now on, we denote by $u$ the longitudinal momentum fraction
of the constituent quark in the emitted meson $M_2$, which can be
factored out from the $(B,M_1)$ system, and by $v$ the momentum
fraction of the quark in the recoiled meson $M_1$, which picks up
the spectator quark from the decaying $B$ meson. For $B$ meson
decaying into two light energetic hadronic final states, we define
the light-cone distribution amplitudes by choosing the $+$
direction along the decay path of the emission meson $M_2$.

Equipped with these necessary preliminaries, the four $B\to\pi K$
and the three $B\to \pi\pi$ decay amplitudes can be expressed
as~\cite{bbns2,bbns3}
\begin{eqnarray} \label{piK}
   {\cal A}(B^-\to\pi^-\overline K^0)
   &=& \lambda_p \left[ \left( a_4^p - \frac12\,a_{10}^p \right)
    + r_\chi^K \left( a_6^p - \frac12\,a_8^p \right) \right]
    X^{(B^-\pi^-,\overline K^0)} \,\nonumber\\
    &&\mbox{}+\left[ \lambda_u\, b_2+(\lambda_u+\lambda_c)(b_3+ b_3^{ew})
    \right] X^{(B^-,\pi^-\overline K^0)} \,, \nonumber\\
   \sqrt2\,{\cal A}(B^-\to\pi^0 K^-)
   &=& \left[ \lambda_u\,a_1 + \lambda_p\,(a_4^p + a_{10}^p)
   + \lambda_p\,r_\chi^K\,(a_6^p + a_8^p) \right]
    X^{(B^-\pi^0,K^-)} \nonumber\\
   && + \left[ \lambda_u\,a_2 + \lambda_p \,\frac32\,(- a_7^p + a_9^p) \right]
    X^{(B^-K^-,\pi^0)} \,\nonumber\\
   && + \left[ \lambda_u\, b_2+(\lambda_u+\lambda_c)(b_3+ b_3^{ew})
    \right] X^{(B^-,\pi^0 K^-)} \,, \nonumber\\
   {\cal A}(\overline B^0\to\pi^+ K^-)
   &=& \left[ \lambda_u\,a_1 + \lambda_p\,(a_4^p + a_{10}^p)
    + \lambda_p\,r_\chi^K\,(a_6^p + a_8^p) \right] X^{(\overline B^0\pi^+,K^-)}
    \, \nonumber\\
    && + (\lambda_u+\lambda_c)\left(b_3-\frac{1}{2}\, b_3^{ew}
       \right) X^{(\overline B^0,\pi^+ K^-)} \,, \nonumber\\
   \sqrt2\,{\cal A}(\overline B^0\to\pi^0 \overline K^0)
   &=& \left[ \lambda_u\,a_2
    + \lambda_p \,\frac32\,(- a_7^p + a_9^p) \right]  X^{(\overline B^0\overline K^0,\pi^0 )}
    \,\nonumber\\
    && - \lambda_p \left[ \left( a_4^p - \frac12\,a_{10}^p \right)
    + r_\chi^K \left( a_6^p - \frac12\,a_8^p \right) \right]
    X^{(\overline B^0\pi^0, \overline K^0)} \,\nonumber\\
   && -(\lambda_u+\lambda_c)\left(b_3-\frac{1}{2}\, b_3^{ew}
       \right) X^{(\overline B^0,\pi^0 \overline K^0)} \,,\\
   {\cal A}(\overline B^0\to\pi^+\pi^-)
   &=& \left[ \lambda_u'\,a_1 + \lambda_p'\,(a_4^p + a_{10}^p)
    + \lambda_p'\,r_\chi^\pi\,(a_6^p + a_8^p) \right]\,
    X^{(\overline B^0\pi^+, \pi^-)} \, \nonumber\\
   && + \left[ \lambda_u'\,b_1 + (\lambda_u'+\lambda_c') \left(
    b_3 + 2 b_4 - \frac{1}{2}\,b_3^{ew} + \frac{1}{2}\,b_4^{ew}
    \right) \right] X^{(\overline B^0,\pi^+ \pi^-)} \,, \nonumber\\
   \sqrt2\,{\cal A}(B^-\to\pi^-\pi^0)
   &=& \left[ \lambda_u' (a_1 + a_2) + \frac32\,\lambda_p'
    (-a_7^p + r_\chi^\pi\,a_8^p + a_9^p + a_{10}^p) \right]\,
    X^{(B^- \pi^-, \pi^0)} \,, \nonumber\\
   {\cal A}(\bar{B}^0\to\pi^0\pi^0)
   &=& \left[ -\lambda_u'\,a_2 + \lambda_p'\,(a_4^p -\frac12 a_{10}^p)
    + \lambda_p'\,r_\chi^\pi\,(a_6^p -\frac12 a_8^p)\,\right.\nonumber\\
   && - \left. \frac32\,\lambda_p'(-a_7^p + a_9^p) \right]\,
   X^{(\overline B^0\pi^0, \pi^0)} \, \nonumber\\
   && + \left[ \lambda_u'\,b_1 + (\lambda_u'+\lambda_c') \left(
    b_3 + 2 b_4 - \frac{1}{2}\,b_3^{ew} + \frac{1}{2}\,b_4^{ew}
    \right) \right] X^{(\overline B^0,\pi^0 \pi^0)} \,,
   \label{pipi}
\end{eqnarray}
where the ``chirally-enhanced" factor $r_\chi^M=r_\chi^M(\mu)$
associated with the coefficients $a_6$ and $a_8$ is defined by
\begin{equation}
   r_\chi^K(\mu)
   = \frac{2\,m_K^2}{\overline{m}_b(\mu)\,
      (\overline{m}_{u,d}(\mu)+\overline{m}_s(\mu))} \,,\qquad
   r_\chi^\pi(\mu)
   = \frac{2 m_\pi^2}{\overline{m}_b(\mu)\,
      (\overline{m}_u(\mu)+\overline{m}_d(\mu))}\,,
\end{equation}
with $\overline{m}_q(\mu)$ being the current quark mass and
depending on the scale $\mu$. The $CP$-conjugated decay amplitudes
are obtained from the above expressions by just replacing
$\lambda_p^{(\prime)}$ with $\lambda_p^{(\prime)*}$.

In Eq.~(\ref{piK}) and ~(\ref{pipi}), we have defined $X^{(\bar{B}
M_1,M_2)}$ as the factorized amplitude with the meson $M_2$ being
factored out from the $(\bar B,M_1)$ system
\begin{equation}
 X^{(\bar{B}M_1,M_2)}=\langle
 M_2|(\bar{q}_2q_3)_{V-A}|0\rangle\cdot
 \langle M_1|(\bar{q}_1b)_{V-A}|\bar{B} \rangle.
\end{equation}
In term of the decay constant and the transition form factors
defined by~\cite{Beneke:2000wa,formfactor}
\begin{eqnarray}
\langle M(p)|\bar{q}\gamma_{\mu} \gamma_5 q'|0 \rangle &=&-i\,
f_{P} p_{\mu},\\
\langle M(p')|\bar q \, \gamma^\mu b |\bar{B}(p)\rangle &=&
F_+^{\bar B\to M}(q^2)\left[p^\mu+p^{\prime\,\mu}-
\frac{m_B^2-m_M^2}{q^2}\,q^\mu\,\right]\,\nonumber\\
&&\mbox + F_0^{\bar B\to M}(q^2)\,\frac{m_B^2-m_M^2}{q^2}\,q^\mu,
\end{eqnarray}
the factorized amplitude can be written as
\begin{equation}
   X^{(\bar{B} M_1,M_2)}= i\,\frac{G_F}{\sqrt2}\,(m_B^2-m_{M_1}^2)\,
    F_0^{\bar B\to M_1}(m_{M_2}^2)\,f_{M_2} \,,
\end{equation}
where we have combined the factor $\frac{G_F}{\sqrt2}$ in the
effective Hamiltonian. The quantity $X^{(\bar{B},M_1M_2)}$
associated with the annihilation coefficient $b_i$ and $b_i^{ew}$
is given by
\begin{equation}
   X^{(\bar{B}, M_1M_2)} = i\,\frac{G_F}{\sqrt{2}}\,f_B\, f_{M_1}\, f_{M_2}
   \,.
\end{equation}

The parameters $a_i\equiv a_i(M_1 M_2)$ in Eq.~(\ref{piK}) and
(\ref{pipi}) encode all the ``non-factorizable'' corrections up to
next-to-leading order in $\alpha_s$, and are calculable with
perturbative theory. The general form of these coefficients
$a_i^p$ can be written as~\cite{bbns3}
\begin{eqnarray}\label{aip}
   a_i^p(M_1 M_2) &=& \left( C_i + \frac{C_{i\pm 1}}{N_c}
   \right)\, \nonumber\\
  && + \,\frac{C_{i\pm 1}}{N_c}\,\frac{C_F\alpha_s}{4\pi}
   \left[ V_i(M_2) + \frac{4\pi^2}{N_c}\,H_i(M_1 M_2) \right]
   + P_i^p(M_2) \,,
\end{eqnarray}
where $C_F=(N_C^2-1)/(2N_C)$, and $N_C=3$ is the number of colors.
The upper (lower) signs apply when $i$ is odd (even) and the
superscript `$p$' should be omitted for $i=1,2$. The first part in
Eq.~(\ref{aip}) corresponds to the NF results, and the remaining
ones to the corrections up to the next-to-leading order in
$\alpha_s$. The quantities $V_i(M_2)$ account for the one-loop
vertex corrections, $H_i(M_1 M_2)$ for the hard spectator
interactions, and $P_i^p(M_1 M_2)$ for the penguin contractions.
In general, these quantities can be written as the convolution of
the hard-scattering kernels with the meson distribution
amplitudes. Explicit form for these quantities are relegated to
Appendix A.

The parameters $b_i\equiv b_i(M_1M_2)$ in Eq.~(\ref{piK}) and
(\ref{pipi}) correspond to the weak annihilation contributions and
are given as~\cite{bbns3}
\begin{eqnarray}\label{bi}
   b_1 &=& \frac{C_F}{N_c^2}\,C_1 A_1^i \,, \qquad
    b_3 = \frac{C_F}{N_c^2} \Big[ C_3 A_1^i + C_5 (A_3^i+A_3^f)
    + N_c C_6 A_3^f \Big] \,, \\
   b_2 &=& \frac{C_F}{N_c^2}\,C_2 A_1^i \,, \qquad
    b_4 = \frac{C_F}{N_c^2}\,\Big[ C_4 A_1^i + C_6 A_2^i \Big] \,, \\
   b_3^{ew} &=& \frac{C_F}{N_c^2} \Big[ C_9 A_1^i
    + C_7 (A_3^i+A_3^f) + N_c C_8 A_3^f \Big] \,, \\
   b_4^{ew} &=& \frac{C_F}{N_c^2}\,\Big[ C_{10} A_1^i + C_8 A_2^i \Big] \,,
\end{eqnarray}
where we have omitted the argument ``$M_1 M_2$". These
coefficients correspond to the current--current annihilation
($b_1,b_2$), the penguin annihilation ($b_3,b_4$), and the
electro-weak penguin annihilation ($b_3^{ew},b_4^{ew}$),
respectively. The explicit form for the building blocks
$A_k^{i,j}$ can be found in Appendix A.

It should be noted that within the QCDF framework, all the
nonfactorizable power suppressed contributions except for the hard
spectator and the annihilation contributions are neglected. We
have re-derived the above next-to-leading order formulas
calculated by Beneke and Neubert~\cite{bbns3}, for which no
deviation has been found.

\subsection{The $b\to D g^\ast g^\ast$ strong penguin
contributions to the $B\to \pi K,\pi\pi$ decays }

From the previous subsection, we can see that, up to
next-to-leading order in $\alpha_s$ and to leading power in
$\Lambda_{\rm QCD}/m_b$, the strong-interaction phases originate
from the imaginary parts of the functions $g(u)$ and $G(s,u)$, as
defined in Eq.~({\ref{vertexfunction}) and Eq.~(\ref{Gfunction}),
respectively. The presence of strong-interaction phase in the
penguin function $G(s,u)$ is well known and commonly referred to
the Bander--Silverman--Soni~(BSS) mechanism~\cite{BSS}. The
reliable calculation of the imaginary part of  function $g(u)$
arising from the hard gluon exchanging between the two outgoing
mesons is a new product of the QCDF approach. However, recent
experimental data indicate that there may exist extra new strong
interaction phases in hadronic $B$-meson decays. Since the $b \to
s g g$ transitions play an important role in the inclusive and
semi-inclusive $B$-meson decays as discussed in literature
\cite{hou1,hou2,simma,greub,bosch,yang:phiXs,mishima}, in this
section we shall generalize these results to exclusive two-body
hadronic $B$ decays, and investigate these $b \to D g^\ast g^\ast$
strong penguin contributions to $B\to \pi K,\pi\pi$ decays.

At the quark level, the $b \to D g^\ast g^\ast$ transitions can
occur in many different manners as depicted by
Figs.\ref{fig:factorizable}--\ref{penguinfig}. For example, one of
the gluons can radiate from the external quark line, while the
other one coming from the chromo-magnetic dipole operator $O_{8g}$
as in Figs.~\ref{Q8gfig}(b) and~\ref{Q8gfig}(c) or from the
internal quark loop in the QCD penguin diagrams in
Figs.~\ref{penguinfig}(b) and~\ref{penguinfig}(c). On the other
hand, the two gluons can also radiate from the internal quark
loops in Figs.~\ref{penguinfig}(d) and~\ref{penguinfig}(e) or
split off the virtual gluon of the strong penguin processes as
shown by Figs.~\ref{Q8gfig}(a) and~\ref{penguinfig}(a). Here we do
not consider the diagrams of the category in
Fig.~\ref{fig:factorizable}, since their contributions can be
absorbed into the definition of the $B\to M_1$ transition form
factors Figs.~\ref{fig:factorizable}(a)
and~\ref{fig:factorizable}(b) or further suppressed by
$\frac{g_s^2}{16\pi^2}$. It is easy to clarify this point by
comparing the strengths of Fig.~\ref{fig:factorizable}(c) to that
of Fig.~\ref{Q8gfig}(a).
\begin{figure}[ht]
\begin{center}
\scalebox{0.7}{\epsfig{file=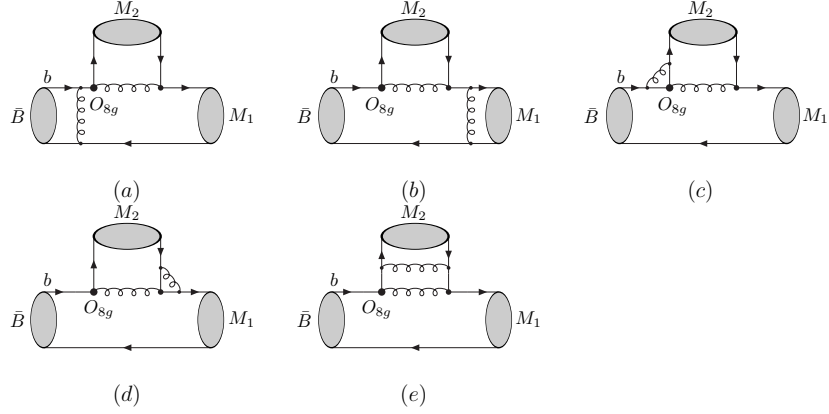}}
\caption{\label{fig:factorizable} \small Representative diagrams
induced by $b\to D g^\ast g^\ast$ transition which are not
evaluated.  Here we give only the chromo-magnetic dipole operator
$Q_{8g}$ contributions. With $\mathcal{O}_{8g}$ replaced by the
other  operators, the corresponding diagrams for these operators
can also be obtained.}
\end{center}
\end{figure}
\begin{figure}[ht]
\begin{center}
\scalebox{0.7}{\epsfig{file=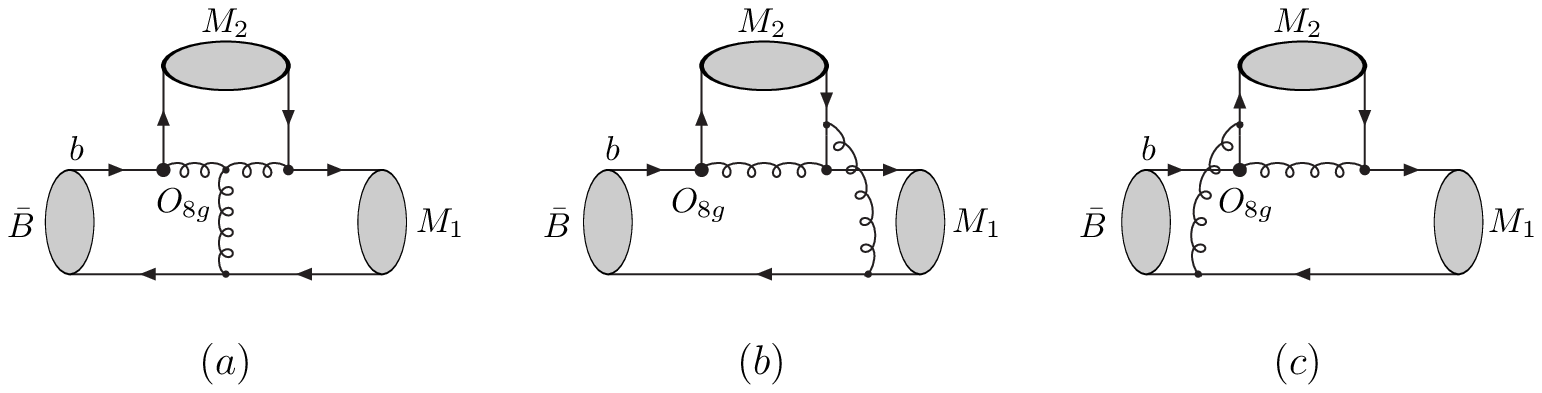}} \caption{\label{Q8gfig}
\small Chromo-magnetic operator $Q_{8g}$ contributions induced by
$b\to D g^\ast g^\ast$ transition.}
\end{center}
\end{figure}
\begin{figure}[ht]
\begin{center}
\scalebox{0.8}{\epsfig{file=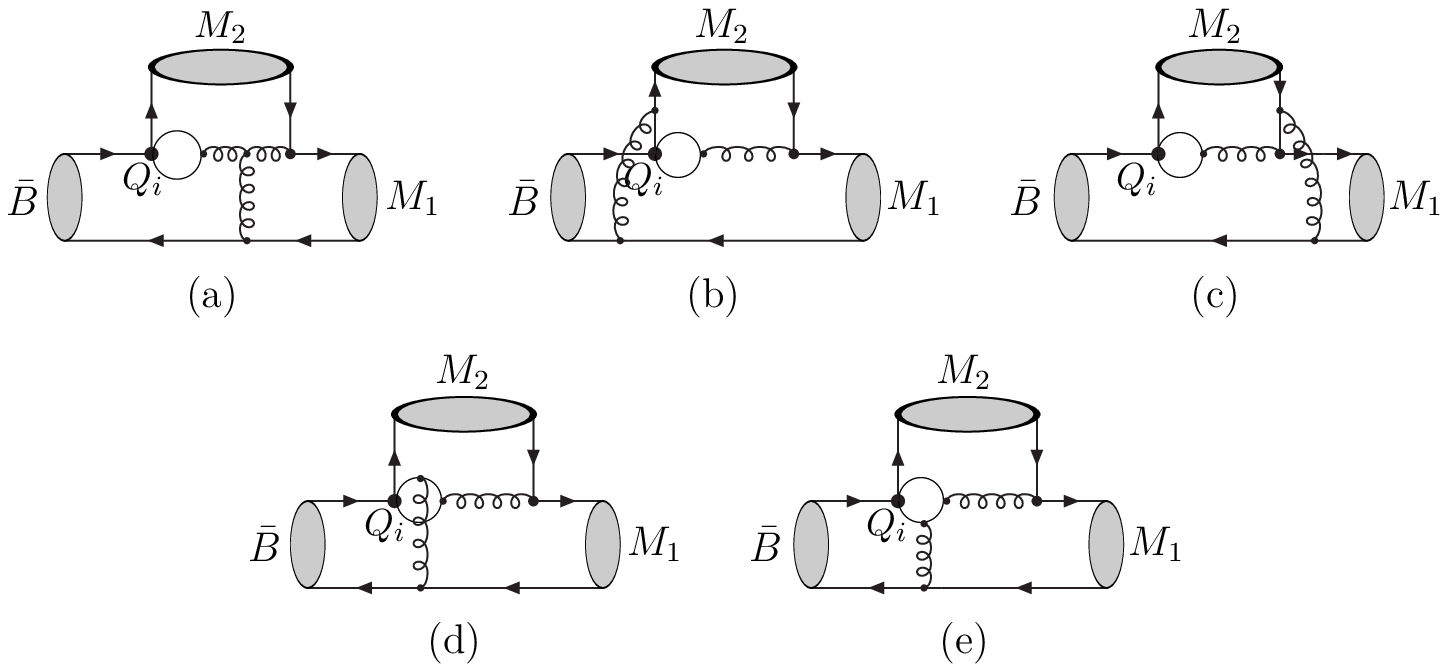}}
\caption{\label{penguinfig} \small Strong penguin contributions
induced by $b\to D g^\ast g^\ast$ transition.}
\end{center}
\end{figure}

As can be seen from Figs.~\ref{Q8gfig} and~\ref{penguinfig}, these
penguin diagrams should be the dominant contributions of order
$\alpha_s^2$, \textit{since they are not two-loop QCD diagrams and
there is no additional $\frac{1}{16 \pi^2}$ suppression factor
compared to the genuine two-loop contributions of order
$\alpha_s^2$}. Studies of these contributions could be helpful for
understanding the higher order perturbative corrections within the
QCDF formalism. In followings,  we first discuss these higher
order strong penguin contributions to decay modes with two light
pseudoscalar mesons in the final states, $B\to M_1M_2$, and then
specialize this general case to the $B\to \pi K,\pi\pi$ decays and
investigate the effect of these higher order corrections on the
branching ratios and $CP$ asymmetries for these modes.

We start with the calculation of the diagrams in
Fig.~\ref{Q8gfig}. In this case, the weak decay is induced by the
chromo-magnetic dipole operator $O_{8g}$. The calculation is
straightforward with the result given by
\begin{eqnarray}
{\cal A}_{Q_{8g}}&=& -i\,\frac{\alpha_s^2\, f_B \,f_{M_1}\,
f_{M_2}}{N_c^3}\,\lambda_t^{(\prime)}\,\int_0^1\!
d{\xi}\,\frac{\Phi_1^B
(\xi)}{\xi}\,\nonumber\\
&&\mbox\,\times\int_0^1\!dudv\,\left\{
\Phi_{M_2}(u)\,\Phi_{M_1}(v)\,\left[\frac{1}{6\,\left( 1 - u
\right) \, \left( 1 - v \right) } +\frac{3\,\left( 3 - v \right) }
{2\,\left( 1 - u \right) \,\left( 1 - v \right) \,v}\,\right]\,\right.\nonumber \\
&&\mbox\,+\left.
r_\chi^{M_1}\,\Phi_{M_2}(u)\,\Phi_p^{M_1}(v)\,\left[\frac{ 2 -
u}{6\,\left(1 - u \right) \,u\,\left(1-v\right)} +\frac{
3\,\left(3-u-v+u\,v \right)}{ 2\,\left(1 - u
\right)^2\,\left(1 - v \right)\,v}\,\right]\,\right.\nonumber\\
&&\mbox\,+\left.
r_\chi^{M_2}\,\Phi_p^{M_2}(u)\,\Phi_{M_1}(v)\,\left[\frac{ 1 +
u}{6\,\left( 1 - u \right) \,\left( 1 - v \right) }
+\frac{3\,\left( 3 - u - v - u\,v \right) } {2\,\left( 1 - u
\right) \,\left( 1 - v \right) \,v}\,\right]\,\right.\nonumber\\
&&\mbox\,+\left.
r_\chi^{M_1}\,r_\chi^{M_2}\,\Phi_p^{M_2}(u)\,\Phi_p^{M_1}(v)\,\left[\frac{
1}{6\,\left(1 - u \right) \,\left( 1 - v \right) }
+\frac{3\,\left( 3 - v \right) } {2\,\left( 1 - u \right) \,\left(
1 - v \right) \,v}\,\right]\,\right\},
\end{eqnarray}
where $\lambda_t=V_{tb}V_{ts}^\ast$ (for $b\to s$ transition) and
$\lambda_t^\prime=V_{tb}V_{td}^\ast$ (for $b\to d$ transition) are
products of the CKM matrix elements. As always, $\Phi_M$ and
$\Phi_p^M$ denote the leading-twist and twist-3 LCDAs of the
pseudoscalar meson $M$ in the final state, respectively.

In calculation of the Feynman diagrams of Fig.~\ref{penguinfig},
we follow the method proposed by Greub and Liniger~\cite{greub}.
First, we calculate the fermion loops in these individual
diagrams, and then insert these building blocks into the entire
diagrams to obtain the total contributions. In evaluating the
internal quark loop diagrams, we shall adopt the naive dimensional
regularization~(NDR) scheme and the modified minimal
subtraction~($\overline{\rm MS}$) scheme. In addition, we shall
adopt the \textit{ad hoc} Feynman gauge throughout this paper.
Similar to the calculation of the penguin contractions in Appendix
A, we should consider the two distinct contractions in the weak
interaction vertex of these penguin diagrams.

As can be seen from Fig.~\ref{penguinfig}, the first three
diagrams have the same building block $I_\mu^a(k)$ (corresponding
to the contraction of operators $Q_{1,3}$) or $\tilde{I}_\mu^a(k)$
(associated with the contractions of the operators $Q_{4,6}$).
These building blocks are shown in Fig.~\ref{buildingblock1} and
given by
\begin{eqnarray}
I_\mu^a(k) &=& \frac{g_s}{4\,\pi^2}\,\Gamma(\frac{\epsilon}{2})\,
(2-\epsilon)\,(4\pi\mu^2)^{\frac{\epsilon}{2}}\,(k_\mu\,\kslash-k^2
\gamma_\mu)\,(1-\gamma_5)\,T^a\,\nonumber\\
&& \mbox\,
\times\int_0^1\!dx\,
\frac{x\,(1-x)}{\left[m_q^2-x(1-x)\,k^2-i\,\delta\,\right]^{\frac{\epsilon}{2}}}\,,\\
\tilde{I}_\mu^a(k) &=&
\frac{g_s}{2\,\pi^2}\,\Gamma(\frac{\epsilon}{2})\,
(4\pi\mu^2)^{\frac{\epsilon}{2}}\,(k_\mu\,\kslash-k^2\gamma_\mu)\,
(1-\gamma_5)\,T^a\,\nonumber\\
&& \mbox\, \times\int_0^1\!dx\,
\frac{x\,(1-x)}{\left[m_q^2-x(1-x)\,k^2-i\,\delta\,\right]^{\frac{\epsilon}{2}}}\,,
\end{eqnarray}
where $k$ and $T^a$ is the momentum and the color generator of the
off-shell gluon, $g_s$ is the strong coupling constant, and $m_q$
the pole mass of the quark propagating in the quark loops. The
free indices $\mu$ and $a$ should be contracted with the gluon
propagator when inserting these building blocks into the entire
diagrams. Here we have used the $d$ dimension space-time as
$d=4-\epsilon$. After performing the subtraction with the
$\overline{\rm MS}$ scheme, we get
\begin{eqnarray}\label{Ibuilding}
I_\mu^a(k) &=&
\frac{g_s}{8\,\pi^2}\left[-\frac{2}{3}-\frac{4}{3}\, \ln
\frac{m_b}{\mu}+G(s_q,1-u)\,\right] (k_\mu\,\kslash-k^2
\gamma_\mu)\,(1-\gamma_5)\,T^a\,,\\
\tilde{I}_\mu^a(k) &=& \frac{g_s}{8\,\pi^2}\left[-\frac{4}{3}\,
\ln \frac{m_b}{\mu}+G(s_q,1-u)\,\right] (k_\mu\,\kslash-k^2
\gamma_\mu)\,(1-\gamma_5)\,T^a\,,
\end{eqnarray}
with the function $G(s,u)$ defined by Eq.~(\ref{Gfunction}).

\begin{figure}[t]
\begin{center}
\scalebox{0.7}{\epsfig{file=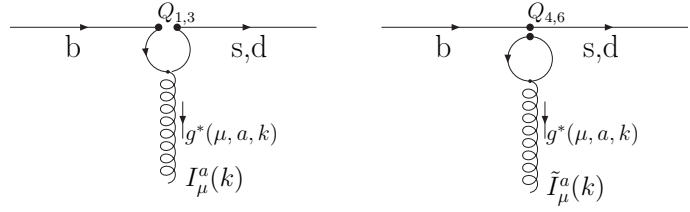}}
\caption{\label{buildingblock1} \small Building blocks
$I_\mu^a(k)$ (associated with the contraction of the operators
$Q_{1,3}$) and $\tilde{I}_\mu^a(k)$ (associated with the
contraction of the operators $Q_{4,6}$) for
Figs.~\ref{penguinfig}(a)--\ref{penguinfig}(c).}
\end{center}
\end{figure}

The sum of the fermion loops in the last two diagrams in
Fig.~\ref{penguinfig} are denoted by the building block
$J_{\mu\nu}^{ab}(k,p)$ (corresponding to the contraction of
operators $Q_{1,3}$) or $\tilde{J}_{\mu\nu}^{ab}(k,p)$
(corresponding to the contraction of operators $Q_{4,6}$), as
depicted by Fig.~\ref{buildingblock2}. Using the decomposition
advocated by \cite{simma,greub}, these building blocks can be
expressed as
\begin{eqnarray}
J^{ab}_{\mu\nu}(k,p) &=& T^{+}_{\mu\,\nu}(k,p)\,\Big\{ T^{a},
T^{b}\Big\} +T^{-}_{\mu\,\nu}(k,p)\Big[ T^{a}, T^{b} \Big]\,,\\
\tilde{J}^{ab}_{\mu\nu}(k,p) &=&
\tilde{T}^{+}_{\mu\,\nu}(k,p)\,\Big\{ T^{a}, T^{b}\Big\}
+\tilde{T}^{-}_{\mu\,\nu}(k,p)\Big[ T^{a}, T^{b} \Big]\,,
\end{eqnarray}
where the first part is symmetric, while the second one is
antisymmetric with respect to the color structures of the two
gluons. Here $k(p)$, $a(b)$, and $\mu(\nu)$ are the momentum,
color, and polarization of the off-shell gluons. Below, we refer
to the gluon with indices $(\nu,b,p)$ as the one connecting with
the spectator quark from the $B$ meson.

\begin{figure}[ht]
\begin{center}
\scalebox{0.55}{\epsfig{file=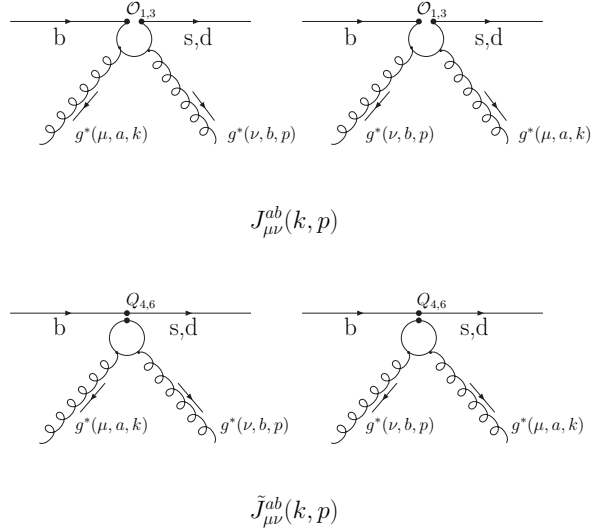}}
\caption{\label{buildingblock2} \small Building blocks
$J_{\mu\nu}^{ab}(k,p)$ (associated with operators $Q_{1,3}$) and
$\tilde{J}_{\mu\nu}^{ab}(k,p)$ (associated with operators
$Q_{4,6}$) for Figs.~\ref{penguinfig}(d) and~\ref{penguinfig}(e).}
\end{center}
\end{figure}

In the NDR scheme, after integrating over the (shifted) loop
momentum, we can present the quantities $T^{\pm}_{\mu\,\nu}(k,p)$
and $\tilde{T}^{\pm}_{\mu\,\nu}(k,p)$ as~\cite{simma,greub}
\begin{eqnarray}\label{TPfunction}
T^{+}_{\mu\,\nu}(k,p)&=& \frac{\alpha_s}{4\,\pi} \,\biggl[
E(\mu,\nu,k)\,\Delta i_{5}\, +E(\mu,\nu, p)\,\Delta i_{6}\,\biggr. \nonumber\\
&&\mbox\,\biggl. -E(\mu, k, p)\, \frac{k_{\nu}}{k\cdot p}\,\Delta
i_{23} - E(\mu, k, p)\, \frac{p_{\nu}}{k\cdot p}\,\Delta i_{24} \biggr. \nonumber\\
&&\mbox\,\biggl. -E(\nu, k, p)\, \frac{k_{\mu}}{k\cdot p}\, \Delta
i_{25} -E(\nu, k, p)\, \frac{p_{\mu}}{k\cdot p}\, \Delta i_{26} \,
\biggr]\, (1-\gamma_5)\,,\\
T^{-}_{\mu\,\nu}(k,p)&=&\label{TMfunction}
\frac{\alpha_s}{4\,\pi}\, \biggl[
\kslash \,g_{\mu\nu}\, \Delta i_{2} +\pslash \,g_{\mu\nu}\, \Delta i_{3} \biggr. \nonumber\\
&&\mbox\,\biggl. +\gamma_{\mu}\, k_{\nu}\, \Delta i_{8}
+\gamma_{\mu}\, p_{\nu}\, \Delta i_{9} +\gamma_{\nu}\, k_{\mu}\,
\Delta i_{11} +\gamma_{\nu}\, p_{\mu}\, \Delta i_{12} \biggr. \nonumber \\
&&\mbox \,\biggl. +\kslash\,\frac{ k_{\mu}k_{\nu} }{k\cdot p}\,
\Delta i_{15} +\kslash\,\frac{ k_{\mu}p_{\nu} }{k\cdot p}\, \Delta
i_{16} +\kslash\,\frac{ p_{\mu}k_{\nu} }{k\cdot p}\, \Delta i_{17}
+\kslash\,\frac{ p_{\mu}p_{\nu} }{k\cdot p}\, \Delta i_{18} \,\biggr. \nonumber \\
&&\mbox \,\biggl. +\pslash\,\frac{ k_{\mu}k_{\nu} }{k\cdot
p}\,\Delta i_{19} +\pslash\,\frac{ k_{\mu}p_{\nu} }{k\cdot p}\,
\Delta i_{20} +\pslash\,\frac{ p_{\mu}k_{\nu} }{k\cdot p}\, \Delta
i_{21} +\pslash\,\frac{ p_{\mu}p_{\nu} }{k\cdot p}\, \Delta i_{22}
\, \biggr]\,(1-\gamma_5)\,,\\
\tilde{T}^{+}_{\mu\,\nu}(k,p)&=& \label{TPtfunction}
a\, T^{+}_{\mu\,\nu}(k,p)\,,\\
\tilde{T}^{-}_{\mu\,\nu}(k,p)&=& T^{-}_{\mu\,\nu}(k,p)\,
+\frac{\alpha_s}{4\,\pi}\, \biggl[ \kslash \,g_{\mu\nu}\,\frac43\,
-\pslash \,g_{\mu\nu}\, \frac43\, -\gamma_{\mu}\, k_{\nu}\,
\frac83\, -\gamma_{\mu}\, p_{\nu}\,\frac43\, \biggr. \nonumber \\
&&\mbox \,\biggl. +\gamma_{\nu}\, k_{\mu}\, \frac43\,
+\gamma_{\nu}\, p_{\mu}\, \frac83\, \biggr]\,(1-\gamma_5)\,,
\end{eqnarray}
where the matrix $E$ in Eq.~(\ref{TPfunction}) is defined by
\begin{eqnarray}
E(\mu,\nu,k) &=& \gamma_\mu \gamma_\nu \kslash\,-\gamma_\mu
k_\nu\,+\gamma_\nu k_\mu\,
-\kslash \, g_{\mu\,\nu}\,\nonumber\\
&=&
-i\,\epsilon_{\mu\nu\alpha\beta}\,k^\alpha\gamma^\beta\gamma_5\,,
\end{eqnarray}
with the second line obtained in a four dimension context with the
Bjorken-Drell conventions. The parameter $a$ in
Eq.~(\ref{TPtfunction}) denotes the chiral structure of the local
four-quark operators in the weak interaction vertex with $a=\pm$
corresponding to $(V-A) \otimes (V\mp A)$, respectively. The
dimensionally regularized expressions for the $\Delta i$ functions
are collected in Appendix B.

Equipped with the explicit form for these building blocks, we can
now evaluate all the Feynman diagrams in Fig.~\ref{penguinfig}.
After direct calculations, the final results with the subscript
denoting the contraction of the corresponding operator in the weak
interaction vertex are
\begin{eqnarray}\label{AQ1}
{\cal A}_{Q_{1}}&=& i\,\frac{\alpha_s^2\, f_B\, f_{M_1}\,
f_{M_2}}{N_c^3}\,\lambda_p^{(\prime)}\,\int_0^1\!
d{\xi}\,\frac{\Phi_1^B(\xi)}{\xi}\,\,\nonumber\\
&&\mbox\,\times \int_0^1\!dudv\,\left[-\frac{2}{3}-\frac{4}{3}\,
\ln{\frac{m_b}{\mu}}+G(s_p,1-u)\,\right]\,f_1(u,v)\,\nonumber\\
&+& i\,\frac{\alpha_s^2\, f_B\, f_{M_1}\,
f_{M_2}}{N_c^3}\,\lambda_p^{(\prime)}\,\int_0^1\!d{\xi}\,\frac{\Phi_1^B(\xi)}{\xi}\,
\int_0^1\!dudv\,f_2(u,v,m_p)\,,\\
{\cal A}_{Q_{3}} &=& -i\,\frac{\alpha_s^2\, f_B\, f_{M_1}\,
f_{M_2}}{N_c^3}\,\lambda_t^{(\prime)}\,\int_0^1\!
d{\xi}\,\frac{\Phi_1^B(\xi)}{\xi}\,\nonumber\\
&&\mbox\,\times \int_0^1\!dudv\,\left[-\frac{4}{3}-\frac{8}{3}\,
\ln{\frac{m_b}{\mu}}+G(0,1-u)+G(1,1-u)\,\right]\,f_1(u,v)\,\nonumber\\
&-& i\,\frac{\alpha_s^2\, f_B\, f_{M_1}\,
f_{M_2}}{N_c^3}\,\lambda_t^{(\prime)}\,\int_0^1\!d{\xi}\,\frac{\Phi_1^B(\xi)}{\xi}\,
\int_0^1\!dudv\,\left[f_2(u,v,0)+f_2(u,v,m_b)\,\right]\,,\\
{\cal A}_{Q_{4}} &=& -i\,\frac{\alpha_s^2\, f_B\, f_{M_1}\,
f_{M_2}}{N_c^3}\,\lambda_t^{(\prime)}\,\int_0^1\!
d{\xi}\,\frac{\Phi_1^B(\xi)}{\xi}\,\int_0^1\!dudv\,\nonumber\\
&&\mbox\,\times \left[-\frac{4\,n_f}{3}\, \ln
\frac{m_b}{\mu}+(n_f-2)\,G(0,1-u)\,+G(s_c,1-u)\,+G(1,1-u)\,\right]\, f_1(u,v)\,\nonumber\\
&-& i\,\frac{\alpha_s^2\, f_B\, f_{M_1}\,
f_{M_2}}{N_c^3}\,\lambda_t^{(\prime)}\,\int_0^1\!
d{\xi}\,\frac{\Phi_1^B(\xi)}{\xi}\,\int_0^1\!dudv\,\nonumber\\
&&\mbox\,\times
\left[(n_f-2)\,f_3(u,v,0)\,+f_3(u,v,m_c)+f_3(u,v,m_b)\,\right]\,,\\
 {\cal A}_{Q_{6}} &=& -i\,\frac{\alpha_s^2\,
f_B\, f_{M_1}\, f_{M_2}}{N_c^3}\,\lambda_t^{(\prime)}\,\int_0^1\!
d{\xi}\,\frac{\Phi_1^B(\xi)}{\xi}\,\int_0^1\!dudv\,\nonumber\\
&&\mbox\,\times \left[-\frac{4\,n_f}{3}\, \ln
\frac{m_b}{\mu}+(n_f-2)\,G(0,1-u)\,+G(s_c,1-u)+G(1,1-u)\right]\,f_1(u,v)\,\nonumber\\
&-& i\,\frac{\alpha_s^2\, f_B\, f_{M_1}\,
f_{M_2}}{N_c^3}\,\lambda_t^{(\prime)}\,\int_0^1\!
d{\xi}\,\frac{\Phi_1^B(\xi)}{\xi}\,\int_0^1\!dudv\,\nonumber\\
&&\mbox\,\times
\left[(n_f-2)\,f_4(u,v,0)\,+f_4(u,v,m_c)+f_4(u,v,m_b)\,\right]\,,
\end{eqnarray}
with
\begin{eqnarray}
f_1(u,v) & = & \Phi_{M_2}(u)\,\Phi_{M_1}(v)\,\left[
\frac{1}{12\,\left( 1 - u \right) \,\left( 1 - v \right) } +
\frac{3\,\left( 3 - 2\,u -v
\right) }{4\,\left( 1 - u \right) \,\left( 1 - v \right) \,v}\,\right]\,\nonumber \\
&+& r_\chi^{M_1}\,\Phi_{M_2}(u)\,\Phi_p^{M_1}(v)\,\left[
\frac{3\,\left( 3 - v \right) }{4\,\left( 1 - u \right) \,\left( 1
- v \right) \,v} + \frac{2 - u}{12\,\left( 1 - u \right) \,u\,
\left( 1 - v \right) }\,\right]\,\nonumber\\
&-&
r_\chi^{M_1}r_\chi^{M_2}\,\Phi_p^{M_2}(u)\,\Phi_p^{M_1}(v)\,\left[
\frac{1}{12\,\left( 1 - u \right) \,\left( 1 - v \right) } -
\frac{3\,\left( 3 - 2\,u - v + 2\,u\,v \right)}{4\,\left( 1 - u
\right) \,\left( 1 - v \right)\,v}\,\right]\,\nonumber\\
&+& r_\chi^{M_2}\,\Phi_p^{M_2}(u)\,\Phi_{M_1}(v)\,\left[
\frac{1}{12\,\left( 1 - v \right) } +
\frac{3\,\left( 3 - v \right) }{4\,\left( 1 - v \right) \,v}\right]\,,\\
f_2(u,v,m_q) & = & \Phi_{M_2}(u)\,\Phi_{M_1}(v)\,\left[
\frac{3\,\Delta i_2}{8\,\left( 1 - u \right) \,\left( 1 -
v\right)} + \frac{3\,\Delta i_3}{8\,\left( 1 - u \right) \,v}
+ \frac{7\,\Delta i_6}{24\,\left( 1 - u \right) \,v}\,\,\right.\,\nonumber\\
&&\left.\,+ \frac{3\,\Delta i_8}{8\,\left( 1 - v \right)\,v}+
\frac{7\,\Delta i_{23}}{24\,\left( 1 - v \right) \,v}+
\frac{7\,\left( 1 - u + v \right) \,\Delta i_5}{24\,\left( 1 - u
\right) \,\left( 1 - v \right) \,v}\,\right]\,\nonumber \\
&-& r_\chi^{M_1}\,\Phi_{M_2}(u)\,\Phi_p^{M_1}(v)\,\left[
\frac{3}{8\,\left( 1 - u \right) \,v}\,\left(\Delta i_3+\Delta
i_{21}\right) + \frac{3\,\Delta i_{12}}{4\,\left( 1 - u
\right) \,v} \,\right.\,\nonumber\\
&& \left. + \frac{7}{24\,\left( 1 - u \right) \,v}\,\left(\Delta
i_6+\Delta i_{26}\right) + \frac{7\,\Delta i_5}{12\,\left( 1 - u
\right) \,\left( 1 -
v\right) } \,\right.\,\nonumber\\
&& \left. + \frac{3}{8\,\left( 1 - u \right)\,
\left( 1 - v \right) }\,\left(\Delta i_2 - \Delta i_8 + \Delta i_{17}\right)\,\right]\,\nonumber\\
&-& r_\chi^{M_2}\,\Phi_p^{M_2}(u)\,\Phi_{M_1}(v)\,\left[
\frac{3\,\Delta i_2}{8\,\left( 1 - v \right) \,v} +
\frac{7\,\Delta i_{23}}{24\,\left( 1 - v \right) \,v} +
\frac{7\,\Delta i_5}{12\,\left( 1 - v \right) \,v} \,\right.\,\nonumber\\
&& \left. - \frac{3\,\Delta i_8}{8\,\left( 1 - v \right) }\,\right]\,\nonumber\\
&+&
r_\chi^{M_1}r_\chi^{M_2}\,\Phi_p^{M_2}(u)\,\Phi_p^{M_1}(v)\,\left[
\frac{7\,\Delta i_{5}}{12\,\left( 1 - v \right) } -
\frac{3\,\Delta i_{12}}{8\,v} - \frac{7\,u\,\Delta
i_{23}}{12\,\left( 1 - u \right) \,\left( 1 - v
\right) } \,\right.\,\nonumber\\
&& \left. + \frac{3\,u}{8\,\left( 1 - u \right) \,v}\,\left(\Delta
i_3+\Delta i_{21}\right)\,+ \frac{7}{24\,\left( 1 - u \right)
\,v}\,\left(\Delta i_6+\Delta i_{26}\right) \,\right.\,\nonumber\\
&& \left. + \frac{3}{8}\,\left( \frac{1}{\left( 1 - u \right)
\,\left( 1 - v \right) } + \frac{1}{v} \right) \,\left(\Delta
i_2+\Delta i_8+\Delta i_{17}\right)\,\right]\,,\\
f_{3,(4)}(u,v,m_q) & = &
\Phi_{M_2}(u)\,\Phi_{M_1}(v)\,\left[-\frac{\left( 3 - 2\,u - 2\,v
\right) }{2\,\left( 1 - u \right) \,\left( 1 - v \right) \,v}+
\frac{3\,\Delta i_2}{8\,\left( 1 - u \right) \,\left( 1 -
v\right)} + \frac{3\,\Delta i_3}{8\,\left( 1 - u \right) \,v}\,\right.\,\nonumber\\
&&\left.\pm \frac{7\,\left( 1 - u + v \right) \,\Delta
i_5}{24\,\left( 1 - u \right) \,\left( 1 - v \right) \,v}\,\pm
\frac{7\,\Delta i_6}{24\,\left( 1 - u \right) \,v}\,+
\frac{3\,\Delta i_8}{8\,\left( 1 - v \right)\,v} \pm
\frac{7\,\Delta i_{23}}{24\,\left( 1 - v \right) \,v}\,\right]\,\nonumber \\
&-&
r_\chi^{M_1}\,\Phi_{M_2}(u)\,\Phi_p^{M_1}(v)\,\left[\frac{3}{2\,\left(
1 - u \right) \,\left( 1 - v \right) \,v}+ \frac{3}{8\,\left( 1 -
u \right) \,v}\,\left(\Delta i_3+\Delta i_{21}\right)\,\right.\,\nonumber\\
&& \left. \pm \frac{7}{24\,\left( 1 - u \right) \,v}\,\left(\Delta
i_6+\Delta i_{26}\right) + \frac{3\,\Delta i_{12}}{4\,\left( 1 - u
\right) \,v} \pm \frac{7\,\Delta i_5}{12\,\left( 1 - u \right)
\,\left( 1 - v\right) } \,\right.\,\nonumber\\
&& \left. + \frac{3}{8\,\left( 1 - u \right)\,
\left( 1 - v \right) }\,\left(\Delta i_2 - \Delta i_8 + \Delta i_{17}\right)\,\right]\,\nonumber\\
&-&  r_\chi^{M_2}\,\Phi_p^{M_2}(u)\,\Phi_{M_1}(v)\,\left[
\frac{3}{2\,\left( 1 - v \right) \,v} + \frac{3\,\Delta
i_2}{8\,\left( 1 - v \right) \,v} \pm \frac{7\,\Delta
i_{23}}{24\,\left( 1 - v \right) \,v} \,\right.\,\nonumber\\
&& \left.\pm \frac{7\,\Delta i_5}{12\,\left( 1 - v \right) \,v}\, - \frac{3\,\Delta i_8}{8\,\left( 1 - v \right) }\,\right]\,\nonumber\\
&+&
r_\chi^{M_1}r_\chi^{M_2}\,\Phi_p^{M_2}(u)\,\Phi_p^{M_1}(v)\,\left[-
\frac{3 - 2\,u - 2\,v + 2\,u\,v}{2\,\left( 1 - u \right) \,\left(
1 - v \right) \,v} \pm \frac{7\,\Delta i_{5}}{12\,\left( 1 - v
\right) } - \frac{3\,\Delta i_{12}}{8\,v} \,\right.\,\nonumber\\
&& \left. \mp \frac{7\,u\,\Delta i_{23}}{12\,\left( 1 - u \right)
\,\left( 1 - v \right) } + \frac{3\,u}{8\,\left( 1 - u \right)
\,v}\,\left(\Delta i_3+\Delta i_{21}\right)\,\right.\,\nonumber\\
&&\left.\,\pm \frac{7}{24\,\left( 1 - u \right)
\,v}\,\left(\Delta i_6+\Delta i_{26}\right) \,\right.\,\nonumber\\
&& \left. + \frac{3}{8}\,\left( \frac{1}{\left( 1 - u \right)
\,\left( 1 - v \right) } + \frac{1}{v} \right) \,\left(\Delta
i_2+\Delta i_8+\Delta i_{17}\right)\,\right]\,,
\end{eqnarray}
where the argument $m_q$ is the quark mass propagating in the
fermion loops. At this stage, the $\Delta_i$ functions are the
ones that have been performed the Feynman parameter integrals,
whose explicit forms can be found in Appendix B.

With the individual operator contributions given above, the total
contributions of these higher order $b\to D g^\ast g^\ast$ strong
penguin diagrams to the decay amplitudes of $B\to M_1M_1$ modes
can be written as
\begin{eqnarray}
{\cal A}_{b\to D g^\ast g^\ast} = \frac{G_F}{\sqrt
2}\,\left[{C_{8g}^{\rm eff}\,\cal A}_{Q_{8g}}+C_1\,{\cal
A}_{Q_{1}}+C_3\,{\cal A}_{Q_{3}}+C_4\,{\cal A}_{Q_{4}}+C_6\,{\cal
A}_{Q_{6}}\,\right]\,.
\end{eqnarray}
In order to specialize these general results to $B\to \pi
K,\pi\pi$ decays, we just need to replace $M_1$ and $M_2$ with the
corresponding mesons. Explicitly, the $b\to D g^\ast g^\ast$
strong penguin contributions to the decay amplitudes of the four
$B\to \pi K$ and the three $B\to \pi\pi$ decay channels are
\begin{eqnarray}\label{piKprime}
{\cal A}^{\prime}(B^-\to \pi^-\overline K^0) &=& {\cal
A}^{\prime}(\overline B^0 \to\pi^+ K^-) = {\cal A}_{b\to s g^\ast
g^\ast}\,(M_1 \to \pi,M_2 \to K)\,,\nonumber\\
\sqrt2\,{\cal A}^{\prime}(B^-\to \pi^0 K^-) &=& -\sqrt2\,{\cal
A}^{\prime}(\overline B^0\to\pi^0 \overline K^0) = {\cal A}_{b\to
s g^\ast g^\ast}\,(M_1 \to \pi,M_2 \to K)\,,\\
{\cal A}^{\prime}(\overline B^0 \to\pi^+ \pi^-) &=& {\cal
A}^{\prime}(\overline B^0\to\pi^0 \pi^0) = {\cal A}_{b\to d g^\ast
g^\ast}\,(M_1\to \pi,M_2 \to \pi)\,,\nonumber\\
{\cal A}^{\prime}(B^-\to \pi^-\pi^0) &=& 0\,,\label{pipiprime}
\end{eqnarray}
where the superscript `$\prime$' is indicated there to be
distinguished from the next-to-leading order results given by
Eqs.~(\ref{piK}) and (\ref{pipi}). The total decay amplitudes are
then the sum of these two pieces.

With the total decay amplitudes, the branching ratio for $B\to M_1
M_2$ decays reads
\begin{eqnarray}
{\cal B}(B \to M_1 M_2)=\frac{\tau_B \,p_c}{8\,\pi\,
m_B^2}\,\left|{\cal A}(B\to M_1 M_2)+{\cal A}^{'}(B\to M_1
M_2)\,\right|^2 \,\cdot\,S\,,
\end{eqnarray}
where $\tau_B$ is the lifetime of the $B$ meson, $S=1/2$ if $M_1$
and $M_2$ are identical, and $S=1$ otherwise. $p_c$ is the
magnitude of the momentum of the final-state particle $M_{1,2}$ in
the $B$ meson rest frame and given by
\begin{eqnarray}
p_c=\frac{1}{2m_B}\sqrt{\left[m_B^2-(m_{M_1}+m_{M_2})^2\,\right]\,
\left[m_B^2-(m_{M_1}-m_{M_2})^2\,\right]}\,.
\end{eqnarray}

As for the direct $CP$ asymmetries, we use the definition of the
difference of the $\bar B$-meson minus $B$-meson decay rates
divided by their sum. With the branching ratios of the
$CP$-conjugated modes denoted by ${\cal B}(\bar B \to \bar f)$,
the $CP$-averaged branching ratios and the direct $CP$ asymmetries
for $B\to f$ decays can be expressed respectively as
\begin{eqnarray}
\bar{\cal B} &=& \frac12\,\left[{\cal B}(\bar B \to \bar f) +
{\cal B}(B\to f)\,\right]\,,\\
{\cal A}_{CP} &=& \frac{{\cal B}(\bar B \to \bar f) -{\cal B}(B\to
f)} {{\cal B}(\bar B \to \bar f) + {\cal B}(B\to f)}\,.
\end{eqnarray}

\section{Numerical calculation and Discussions}

\subsection{Input parameters}

The theoretical predictions with the QCDF approach depend on many
input parameters such as the CKM matrix elements, Wilson
coefficients, hadronic parameters, and so on. We present all the
relevant input parameters as follows.

\noindent { -\bf Wilson coefficients.} The Wilson coefficients
$C_i(\mu)$ in the effective weak Hamiltonian have been reliably
evaluated to the next-to-leading logarithmic order. To proceed, we
use the following numerical values at $\mu=m_b$ scale, which have
been obtained in the NDR scheme~\cite{Buchalla:1996vs,Buras:2000}
\begin{equation}
\begin{array}{llll}
    C_1= 1.082,&
    C_2= -0.185,&
    C_3=  0.014,&
    C_4= -0.035,\\
    C_5=  0.009,&
    C_6= -0.041,&
    C_7/\alpha= -0.011,&
    C_8/\alpha=  0.059,\\
    C_9/\alpha= -1.241,&
    C_{10}/\alpha= 0.218,&
    C_{7\gamma}^{\rm eff}=-0.299,&
    C_{8g}^{\rm eff}=-0.143.
\end{array}
\end{equation}
\newline
\noindent { -\bf The CKM matrix elements.} The widely used
parametrization of the CKM matrix elements in analyzing $B$-meson
decays is the Wolfenstein parametrization, which emphasizes the
hierarchies among its elements and is expanded as a power series
in the parameter $\lambda =|V_{us}|$~\cite{Wolfenstein:1983yz},
\begin{equation}
V_{\rm CKM}= \left( \begin{array}{ccc}
1-\frac{\lambda^{2}}{2} &  \lambda &  A\lambda^{3}(\rho-i\eta) \\
-\lambda     & 1-\frac{\lambda^{2}}{2}  & A\lambda^{2}   \\
A\lambda^{3}(1-\rho-i\eta)& -A\lambda^{2}    &   1  \\
\end{array}  \right)\,+{\cal O}(\lambda^4)\,.
\end{equation}
The values of the four Wolfenstein parameters ($A$, $\lambda$,
$\rho$, and $\eta$) could be determined from the best knowledge of
the experimental and theoretical inputs. In this paper, we take
\begin{eqnarray}
A=0.8533,\qquad \lambda=0.2200,\qquad \bar{\rho}=0.20,\qquad
\bar{\eta}=0.33,
\end{eqnarray}
as our default input values~\cite{PDG2004}. The parameters $\bar
\rho$ and $\bar \eta$ are defined by
$\bar{\rho}=\rho\,(1-\frac{\lambda^2}{2})\,,
\bar{\eta}=\eta\,(1-\frac{\lambda^2}{2})$.

\noindent { -\bf Masses and lifetimes.} For the quark mass, there
are two different classes appearing in the QCDF approach. One type
is the pole quark mass which appears in the evaluation of the
penguin loop corrections, and is denoted by $m_q$ with
$q=u,d,s,c,b$. In this paper, we take
\begin{eqnarray}
m_u=m_d=m_s=0,\qquad m_c=1.47\,{\rm GeV}, \qquad m_b=4.80\,{\rm
GeV},
\end{eqnarray}
as our default input values.

The other one is the current quark mass which appears in the
equations of motion and is used to calculate the matrix elements
of the penguin operators as well as the chiral enhancement factors
$r_\chi^M$. This kind of quark mass is scale dependent. To get the
corresponding value at the given scale, we should use the
renormalization group equation to run them, which can be found,
for example, in \cite{Buchalla:1996vs}. Following
Ref.~\cite{bbns3}, we hold
$(\overline{m}_u+\overline{m}_d)/\overline{m}_s$ fixed and use
$\overline{m}_s$ as an input parameters. Explicitly, we take
\begin{eqnarray}
\overline{m}_{u}(2\,{\rm GeV})&=&\overline{m}_{d}(2\,{\rm
GeV})=0.0413\,\overline{m}_{s}(2\,{\rm GeV})\,,\nonumber\\
\overline{m}_{s}(2\,{\rm GeV}) &=& 90\,{\rm MeV}\,,\qquad
\overline{m}_{b}(m_b)= 4.40\,{\rm GeV}\,.
\end{eqnarray}
where the difference between the $u$ and $d$ quark is not
distinguished.

For meson masses and the lifetimes of the $B$ meson, we adopt the
center values given by~\cite{PDG2004}
\begin{center}
\begin{tabular}{cccc}
$\tau_{B_{u}}=1.671$\,{\rm ps}\,, & $\tau_{B_{d}}=1.536$\,{\rm
ps}\,,& $ m_{B_{u}}=5.2794$\,{\rm GeV}\,,& $m_{B_{d}}=5.2790$\,{\rm GeV}\,,\\
$ m_{K^{\pm}}=493.7$\,{\rm MeV}\,, & $m_{K^{0}}=497.6$\,{\rm
MeV}\,,& $m_{\pi^{\pm}}=139.6$\,{\rm MeV}\,,&
$m_{\pi^{0}}=135.0$\,{\rm MeV}\,.
\end{tabular}
\end{center}

\noindent { -\bf Light-cone distribution amplitudes of mesons.}
Since the QCDF approach is based on the heavy quark assumption, to
a very good approximation, we can use the asymptotic form of the
LCDAs for light mesons~\cite{Beneke:2000wa,formfactor,Braun:1989iv}
\begin{equation}
\Phi_M (x) = 6\,x(1-x),\qquad \Phi_p^M (x) = 1.
\end{equation}
With respect to the endpoint divergence associated with the
momentum fraction integral over the LCDAs appearing in this paper,
in analogy to the treatment in Refs.~\cite{bbns2,Feldmann}, we
regulate the integral with an \textit{ad-hoc} cut-off
\begin{eqnarray}
\int_0^1 \!dv\,\frac{\Phi_p^M(v)}{1-v}\, &\to&
\int_0^{1-\Lambda_h/m_B} \!dv\,\frac{\Phi_p^M(v)}{1-v} = \ln
\frac{m_B}{\Lambda_h},
 \end{eqnarray}
with $\Lambda_h=500\,{\rm MeV}$, and do not distinguish whether
this divergence comes from the hard spectator rescattering or from
the annihilation contributions. The possible complex phase
associated with this integral has also been neglected.

As for the $B$ meson wave functions, within our approximation, we
need only consider the first inverse moment of the LCDA $\Phi_1^B
(\xi)$ defined by~\cite{bbns2}
\begin{equation}\label{PhiB1}
  \int_0^1 \frac{d\xi}{\xi}\,\Phi_1^B(\xi)
  \equiv \frac{m_B}{\lambda_B}\,,
\end{equation}
where the hadronic parameter $\lambda_B$ has been introduced to
parameterize this integral. This parameter has been evaluated
using different methods~\cite{Braun,Khodjamirian} recently. In
this paper, we take $\lambda_B=460\,{\rm MeV}$ as our default
input value~\cite{Braun}.

\noindent {-\bf Decay constants and transition form factors.} The
decay constant and the form factors are nonperturbative parameters
and can be determined from experiments and/or theoretical
estimations. For the decay constants, we take
\begin{equation}
f_B=200\,{\rm MeV}\,\cite{bbns3},\qquad f_K=160\,{\rm
MeV}\,,\qquad f_\pi=131\,{\rm MeV}.
\end{equation}

For the form factors involving the $B\to K$ and $B\to\pi$
transitions, we take
\begin{equation}
F_0^{B\to \pi}(0) = 0.258\,,\qquad F_0^{B\to K}(0) = 0.331\,,
\end{equation}
as the default values at the maximum recoil. In addition, we use
the formula
\begin{equation}
F_0^{B\to M}(q^2) =\frac{r_2^M}{1-q^2/m_{fit}^2(M)}\,,
\end{equation}
to parameterize the dependence of the form factor on the
momentum-transfer $q^2$, with the fit parameters given by
\begin{equation}
r_2^\pi =0.258\,,\qquad m_{fit}^2(\pi)=33.81\,,\qquad r_2^K
=0.330\,,\qquad m_{fit}^2(K)=37.46\,.
\end{equation}
All of these values are taken from the latest QCD sum rule
analysis~\cite{BallZwicky}.

\subsection{Numerical results and discussions }

With the theoretical expressions and the input parameters given
above, we can now evaluate the branching ratios and the direct
$CP$ asymmetries for $B \to \pi K$ and $B \to \pi \pi$ decays. For
each quantity, we first give the predictions at the
next-to-leading order in $\alpha_s$, and then take into account
the $b \to D g^\ast g^\ast $ strong penguin corrections, which are
of order $\alpha_s^2$. The combining contributions of the two
pieces, denoted by ${\cal O}(\alpha_s+\alpha_s^2)$, is then given
in the last. For comparison, the NF results are also presented.
All the averaged  experimental data are taken from
HFAG~\cite{HFAG}.

\subsubsection{The $CP$-averaged branching ratios for $B \to \pi K,\pi\pi$ decays}

In the SM, the four $B \to \pi K$ decays are dominated by the
$b\to s$ strong penguin diagrams, with additional subdominant
contributions from the tree and electro-weak penguin diagrams. The
three $B\to \pi\pi$ decays, however, are tree-dominated modes. It
is therefore expected that these higher order strong penguin
diagrams considered in this paper should contribute effectively to
$B\to \pi K$ modes, while have only a minor impact on $B\to
\pi\pi$ ones. Numerical results of the $CP$-averaged branching
ratios for these modes are collected in Table~\ref{brs}.

\begin{table}[ht]
\caption{ \label{brs} The $CP$-averaged branching ratios~(in units
of $10^{-6}$) for $B\to \pi K,\pi\pi$ decays with the default
input parameters. $\bar{\cal B}^f$ and $\bar{\cal B}^{f+a}$ denote
the results without and with the annihilation contributions,
respectively. The NF results, which are of order ${\cal
O}(\alpha_s^0)$, are also shown for comparison. $\bar{\rho}=0.20 $
and $\bar{\eta}=0.33 $}
\begin{center}
\doublerulesep 0.8pt \tabcolsep 0.1in
\begin{tabular}{lccccccc}\hline\hline
 \multicolumn{2}{c@{\hspace{-8cm}}}{$\bar{\cal B}^f$} &
 \multicolumn{2}{c@{\hspace{-8cm}}}{$\bar{\cal B}^{f+a}$} \\
\cline{3-4} \cline{5-6}\raisebox{2.3ex}[0pt]{Decay
Mode}&\raisebox{2.3ex}[0pt]{NF}& ${\cal O}(\alpha_s)$ & ${\cal
O}(\alpha_s+\alpha_s^2)$ & ${\cal
O}(\alpha_s)$ & ${\cal O}(\alpha_s+\alpha_s^2)$& \raisebox{2.3ex}[0pt]{Exp.}\\
\hline
 $B^{-} \to \pi^{-} \overline{K}^0 $
 & $10.07$&$13.28$&$17.31$&$16.04$&$20.44$&$24.1 \pm 1.3 $\\
 $B^{-} \to \pi^0 K^{-}$
 & $5.69$&$7.30$&$9.37$&$8.72$&$10.97$&$12.1\pm 0.8 $\\
 $\overline{B}^0 \to \pi^{+} K^{-} $
 & $7.71$&$10.25$&$13.61$&$12.46$&$16.15$& $18.2\pm 0.8$\\
 $\overline{B}^0 \to \pi^0 \overline{K}^0$
 & $3.38$&$4.63$&$6.26$&$5.70$&$7.50$& $11.5\pm 1.0$\\
\hline
 $\overline{B}^0 \to \pi^{+} \pi^{-} $
 & $7.41$&$7.69$&$7.99$&$8.32$&$8.63$& $4.5\pm 0.4$\\
 $B^{-} \to \pi^{-} \pi^0 $
 & $5.12$&$5.06$&$5.06$&$-$&$-$&$5.5\pm 0.6 $\\
 $\overline{B}^0 \to \pi^0 \pi^0$
 & $0.15$&$0.16$&$0.19$&$0.17$&$0.21$& $1.45\pm 0.29$\\
\hline\hline
\end{tabular}
\end{center}
\end{table}

The dependence of these $CP$-averaged branching ratios on the weak
phase $\gamma$ is shown by Fig.~\ref{fig:qcdf}~(without the
annihilation contributions) and Fig.~\ref{fig:annihilation}~(with
the annihilation contributions), where the solid and dashed curves
correspond to the theoretical predictions with and without the
$b\to D g^\ast g^\ast$ strong penguin contributions included,
respectively. The horizontal solid lines denote the experimental
data as given in Table~\ref{brs}, with the thicker one denoting
its center value and the thinner ones its error bars. In these and
the following figures, the default values of all inputs parameters
except for the CKM angle $\gamma$ are used.

\begin{figure}[t]
\begin{center}
\scalebox{0.8}{\epsfig{file=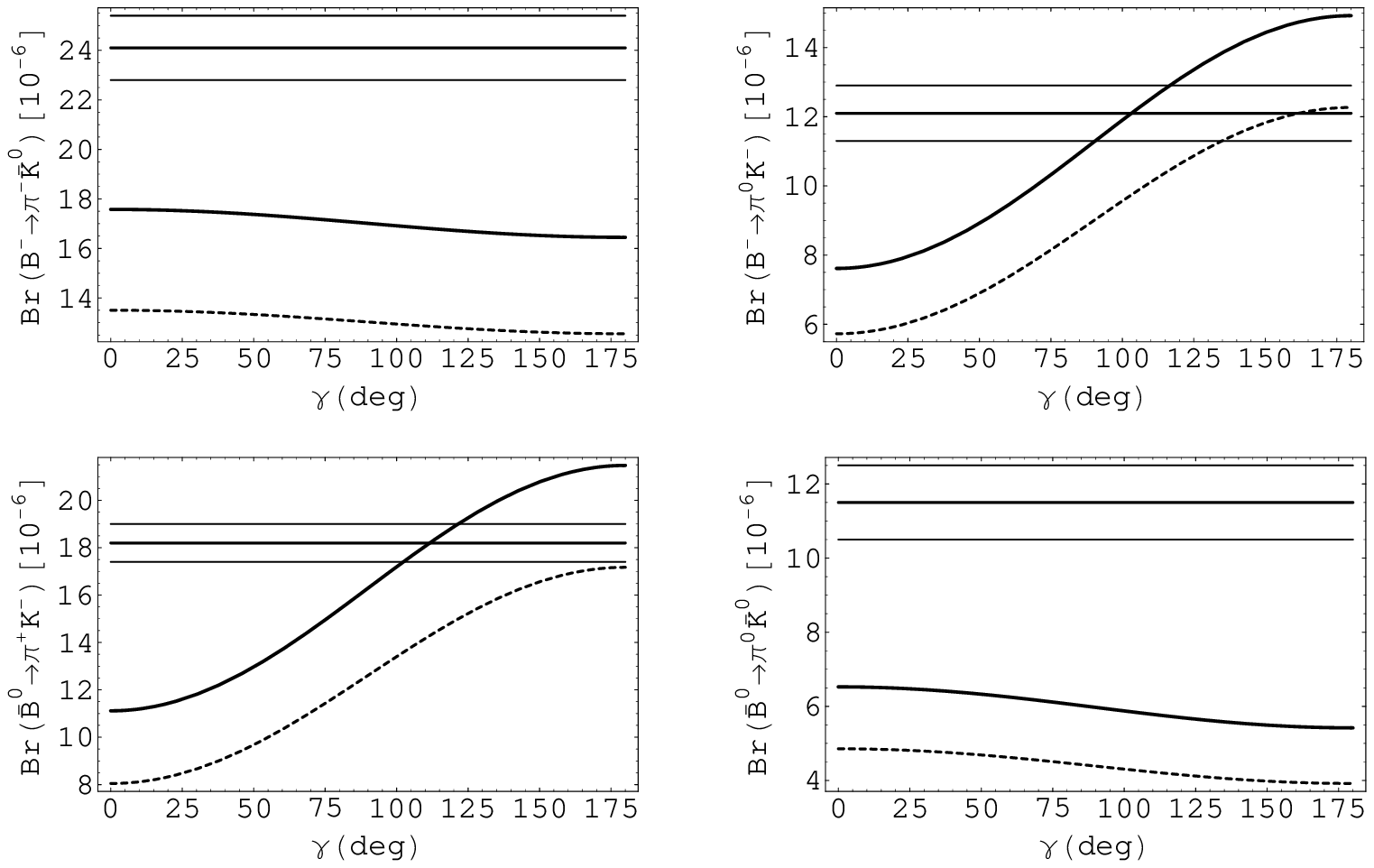}}\\
\scalebox{0.8}{\epsfig{file=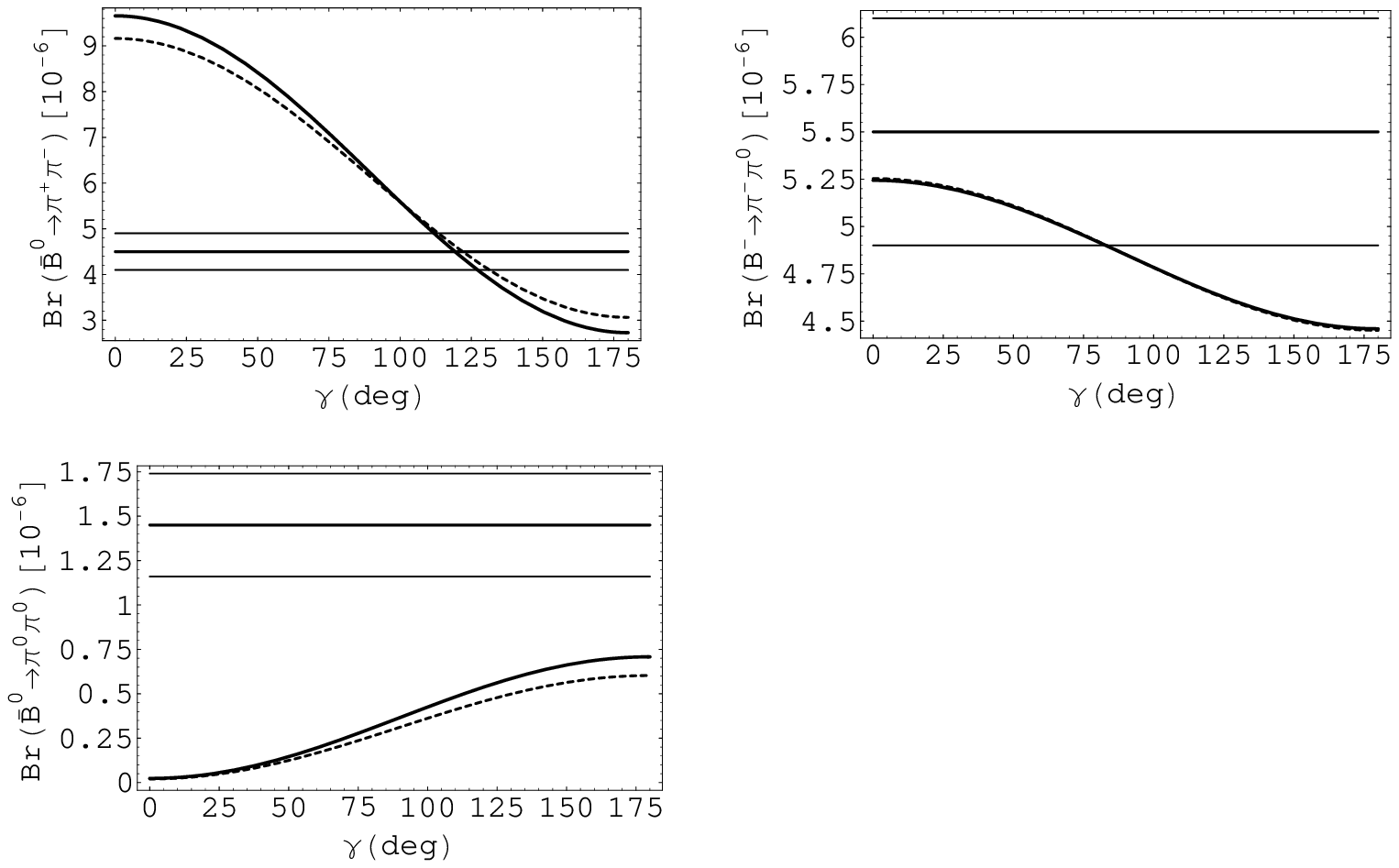}} \caption{The $\gamma$
dependence of the $CP$-averaged branching ratios for $B \to \pi
K,\pi\pi$ decays without the annihilation contributions. The solid
and dashed curves correspond to the theoretical predictions with
and without the $b\to D g^\ast g^\ast$ strong penguin
contributions included, respectively. The horizontal solid lines
denote the experimental data as given in Table~\ref{brs}, with the
thicker ones being its center values and the thinner its error
bars.} \label{fig:qcdf}
\end{center}
\end{figure}

\begin{figure}[t]
\begin{center}
\scalebox{0.8}{\epsfig{file=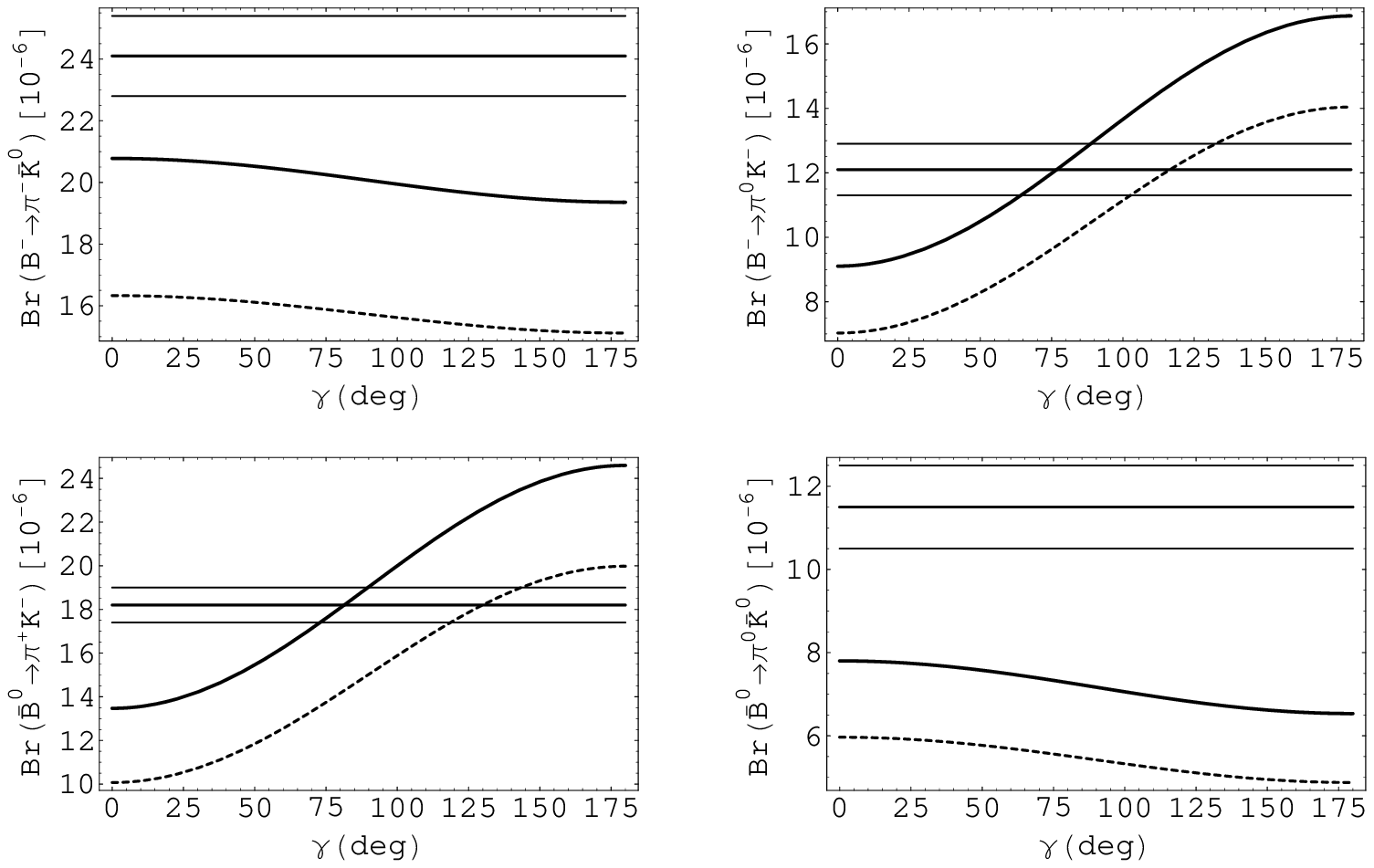}}\\
\scalebox{0.8}{\epsfig{file=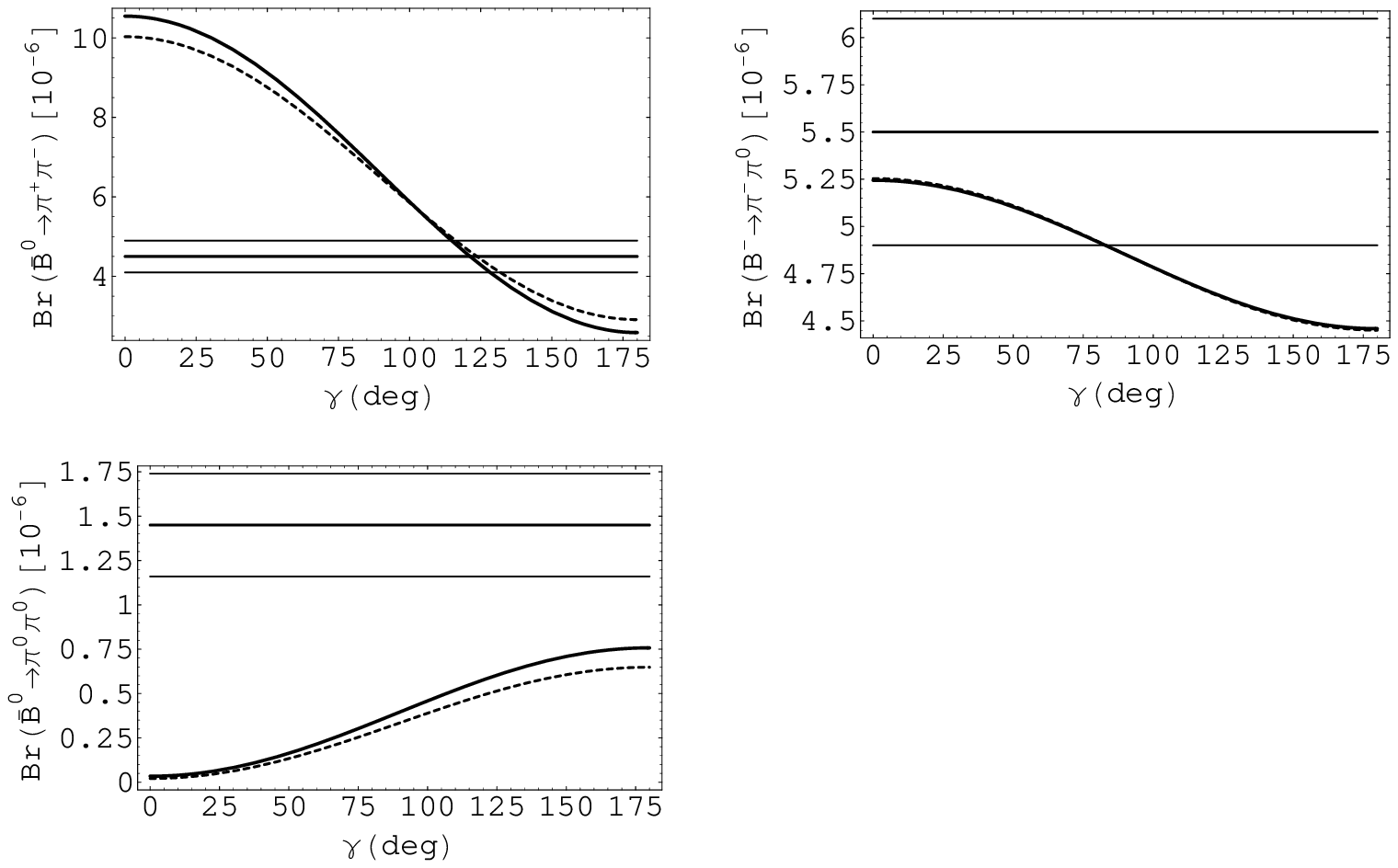}} \caption{The $\gamma$
dependence of the $CP$-averaged branching ratios for $B \to \pi
K,\pi\pi$ decays with the annihilation contributions included. The
meaning of the curves and the horizontal solid lines is the same
as in Fig.~\ref{fig:qcdf}.} \label{fig:annihilation}
\end{center}
\end{figure}

From these two figures and the numerical results given by
Table~\ref{brs}, we can see that:
\begin{itemize}

\item For penguin-dominated $B\to \pi K$ decays, due to the
enhancement of the penguin amplitudes, the QCDF scheme prefers
larger branching ratios than the NF approximation. With our
default input parameters, however, predictions for the branching
rations are still smaller than the experimental data even after
the inclusion of the annihilation contributions, if we consider
only contributions up to the next-to-leading order in $\alpha_s$.
The effects of these higher order $b\to s g^\ast g^\ast$ strong
penguin corrections are very prominent in these penguin-dominated
$B\to \pi K$ decays. With our input parameters, we find that these
higher order strong penguin contributions can give $\sim 30\%$
enhancement to the corresponding branching ratios, and such an
enhancement can improve the consistency between the theoretical
predictions and the experimental data significantly. In addition,
we find that the effect of the annihilation contributions on the
branching ratios, though not negligible, is not so large as
claimed by pQCD method~\cite{pqcd,CDLU}.

\item For tree-dominated $B\to \pi\pi$ decays, the higher order
$b\to d g^\ast g^\ast$ contributions play only a minor role.  To a
very good approximation, the $B^{\pm}\to \pi^{\pm} \pi^0$ decay
can be considered as a pure tree process, and  it does not receive
annihilation contributions either. The theoretical prediction for
the corresponding branching ratio agrees with the data quite well.
For the other two $B\to \pi\pi$ modes, however, theoretical
predictions with QCDF approach are quite inconsistent with the
measured ratios, even with the annihilation and the higher order
strong penguin contributions included. With our input parameters,
we find that the theoretical prediction for $\bar{B}^0 \to \pi^0
\pi^0$ mode is about an eighth of the experimental data; for
$\bar{B}^0 \to \pi^+ \pi^-$ mode, on the other hand, a value about
two times larger than the data is predicted.

\item As for the $\gamma$ dependence of the corresponding
branching ratios, we can see that the two decay modes, $B^{\pm}\to
\pi^{\pm}\pi^0$ and $B^{\pm}\to \pi^{\pm}K^0$, are almost
independent of this angle, since the corresponding decay
amplitudes have to a good approximation only a single weak phase.
In addition, the discrepancy between the theoretical prediction
and the experimental data for $\overline{B}^0 \to \pi^+ \pi^-$ can
be removed if we use a large angle $\gamma \sim 120^\circ$. With
the annihilation and the higher order strong penguin contributions
included, the four $B\to \pi K$ modes, however, prefer a smaller
value for this angle around $\gamma \sim 80^\circ$, which is quite
consistent with the latest direct experimental measurement $\gamma
= 81^\circ \pm 19^\circ(stat.) \pm 13^\circ (sys.) \pm 11^\circ
(model)$~\cite{abe04}.

\item The theoretical predictions for the branching ratios are
very sensitive to the value of the form factor $F_0^{B\to \pi}$.
For example, the large measured decay rates for the four $B\to \pi
K$ decays can be well accommodated with a larger value of the form
factor as shown by Beneke and Neubert~\cite{bbns3}. On the other
hand, the prediction for $\overline{B}^0 \to \pi^+ \pi^-$ decays
can become consistent with the data only when a smaller value is
used. The large measured ratio for $\overline{B}^0 \to \pi^0
\pi^0$, however, remains unresolved with the varying of these
parameters. It is a tough theoretical challenge to accommodate the
current experimental data in the SM.
\end{itemize}

Since the uncertainties in the predictions for  branching ratios
can be largely eliminated by taking  ratios between them, we now
discuss the variations of the quantities defined by
Eqs.~(\ref{Rdefine1})--(\ref{Rdefine5}) with the higher order
$b\to D g^\ast g^\ast$ strong penguin contributions included. It
is the known $``\pi K"$ puzzle~\cite{BF,Buras:2004th} that the SM
predictions are inconsistent with current experiment data.  The
theoretical predictions and the current experimental data for
these ratios are collected in Table~\ref{Rvalue}. For the $\gamma$
dependence of these quantities, we display them in
Fig.~\ref{fig:Rvalue}, where the curves and the horizontal solid
lines have the same interpretations as in Fig.~\ref{fig:qcdf}.

\begin{table}[t]
\caption{ Ratios between the $CP$-averaged branching fractions for
$B\to \pi K,\pi\pi$ modes. The values in the parentheses are the
ones without the annihilation contributions.}
\begin{center}
\doublerulesep 0.8pt \tabcolsep 0.1in
\begin{tabular}{ccccc}\hline\hline
 &NF&${\cal O}(\alpha_s)$ & ${\cal O}(\alpha_s+\alpha_s^2)$ & Exp.\\
 \hline
 $R_{+-}$&$1.272$& $1.119~(1.209)$&$1.077~(1.163)$&$2.20\pm 0.31$\\
 $R_{00}$&$0.040$& $0.041~(0.042)$&$0.048~(0.047)$&$0.67\pm 0.14$\\
 \hline
 $R$&$0.833$& $0.845~(0.840)$&$0.860~(0.855)$&$0.82\pm 0.06$\\
 $R_c$&$1.130$& $1.087~(1.100)$&$1.074~(1.083)$&$1.00\pm 0.09$\\
 $R_n$&$1.140$& $1.092~(1.106)$&$1.077~(1.087)$&$0.79\pm 0.08$\\
 \hline\hline
\end{tabular}\label{Rvalue}
\end{center}
\end{table}

\begin{figure}[t]
\begin{center}
\scalebox{0.7}{\epsfig{file=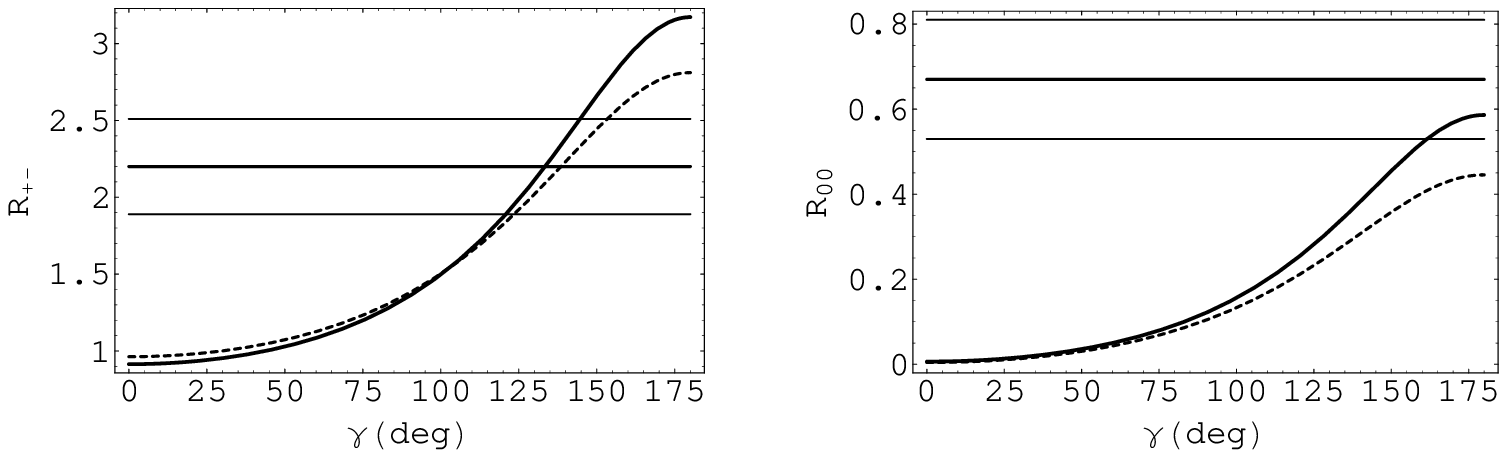}}\\
\scalebox{0.7}{\epsfig{file=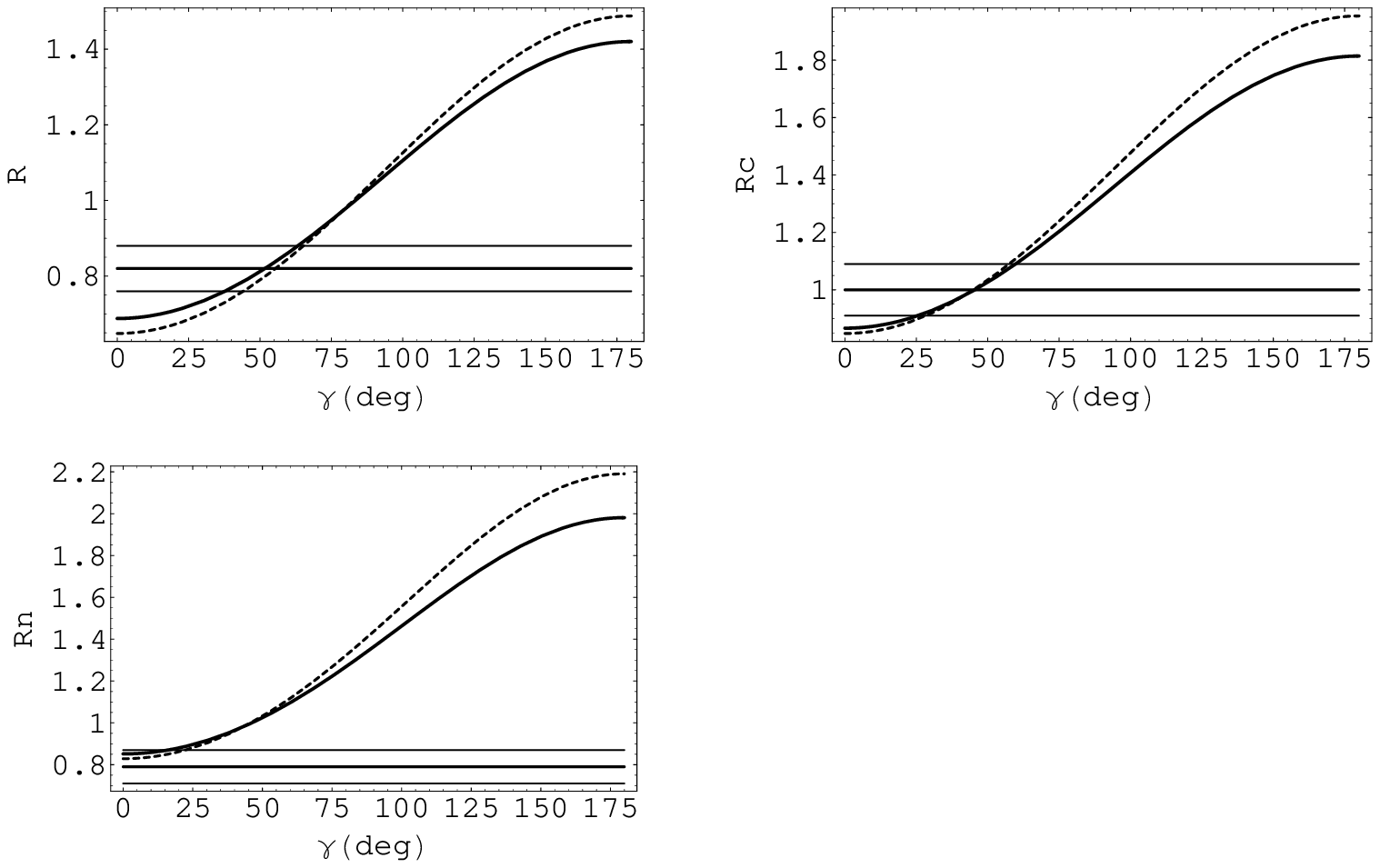}} \caption{
\label{fig:Rvalue} Ratios of the $CP$-averaged branching fractions
defined by Eqs.~(\ref{Rdefine1})--(\ref{Rdefine5}) as functions of
the weak phase $\gamma$. The meaning of the curves and the
horizontal solid lines is the same as in Fig.~\ref{fig:qcdf}.}
\end{center}
\end{figure}

From Table~\ref{Rvalue} and Fig.~\ref{fig:Rvalue}, we can find
that the two ratios $R_c$ and $R_n$ are indeed approximately equal
within the SM as claimed in Ref.~\cite{Buras:2004th}, while the
experimental data for the two quantities are quite different with
the puzzling pattern $R_n<1$. On the other hand, the value of the
quantity $R$ predicted by the QCDF approach is well consistent
with the experimental data. For the other two ratios $R_{+-}$ and
$R_{00}$, the discrepancies between the theoretical predictions
and the experimental data are quite large. As the $b\to D g^\ast
g^\ast$ strong penguin contributions to $B\to \pi K, \pi\pi$
decays are similar in nature, and hence eliminated in the ratios
between the corresponding branching fractions, the patterns of the
these quantities remain unaffected even with these new strong
penguin contributions included. From the $\gamma$ dependence of
the ratios between the four $B\to \pi K$ decays, a smaller value
for this phase is preferred. On the other hand, a larger value for
this phase is favored by $B\to \pi\pi$ decays. These
inconsistences may be hints for new physics playing in the
electroweak penguin sector as suggested  by Buras {\it et
al}.~\cite{Buras:2004th}.

\subsubsection{The direct $CP$ asymmetries for $B \to \pi K, \pi\pi$ decays}

Contrary to the NF approximation, the QCDF scheme can predict the
strong interaction phases and hence the direct $CP$ asymmetries in
the heavy quark limit. The numerical results and the experimental
data for this quantity involving the four $\pi K$ and the three
$\pi\pi$ final states are collected in Table~\ref{ACP}. The
$\gamma$ dependence of the direct $CP$ asymmetries is displayed in
Fig.~\ref{fig:ACP1}~(without the annihilation contributions) and
Fig.~\ref{fig:ACP2}~(with the annihilation contributions), in
which the curves and the horizontal solid lines also have the same
interpretation as in Fig.~\ref{fig:qcdf}.

\begin{table}[t]
\caption{ The direct $CP$ asymmetries~(in units of $10^{-2}$) for
$B\to \pi K,\pi\pi$ decays with the default input parameters.
$A_{CP}^f$ and $A_{CP}^{f+a}$ denote the results without and with
the annihilation contributions, respectively.}
\begin{center}
\doublerulesep 0.8pt \tabcolsep 0.1in
\begin{tabular}{lcccccc}\hline\hline
 \multicolumn{2}{c@{\hspace{-5cm}}}{$A_{CP}^f$} &
 \multicolumn{2}{c@{\hspace{-5cm}}}{$A_{CP}^{f+a}$} \\
\cline{2-3} \cline{4-5}\raisebox{2.3ex}[0pt]{Decay Mode}& ${\cal
O}(\alpha_s)$ & ${\cal O}(\alpha_s+\alpha_s^2)$ & ${\cal
O}(\alpha_s)$ & ${\cal O}(\alpha_s+\alpha_s^2)$& \raisebox{2.3ex}[0pt]{Exp.}\\
\hline
 $B^{-} \to \pi^{-} \overline{K}^0 $
 & $0.73$&$0.52$&$0.65$&$0.46$&$-2.0\pm 3.4 $\\
 $B^{-} \to \pi^0 K^{-}$
 & $7.59$&$6.94$&$6.56$&$6.07$&$4\pm 4 $\\
 $\overline{B}^0 \to \pi^{+} K^{-} $
 & $5.31$&$4.83$&$4.39$&$4.08$& $-10.9\pm 1.9$\\
 $\overline{B}^0 \to \pi^0 \overline{K}^0$
 & $-3.08$&$-2.84$&$-2.71$&$-2.54$& $-9\pm 14$\\
 \hline
 $\overline{B}^0 \to \pi^{+} \pi^{-} $
 &$-4.73$&$-5.51$&$-4.54$&$-5.27$& $37\pm 10$\\
 $B^{-} \to \pi^{-} \pi^0 $
 &$-0.30$&$-0.31$&$-$&$-$&$-2\pm 7$\\
 $\overline{B}^0 \to \pi^0 \pi^0$
 &$55.52$&$58.53$&$55.03$&$55.50$& $28\pm 39$\\
 \hline\hline
\end{tabular}
\end{center}\label{ACP}
\end{table}

\begin{figure}[t]
\begin{center}
\scalebox{0.8}{\epsfig{file=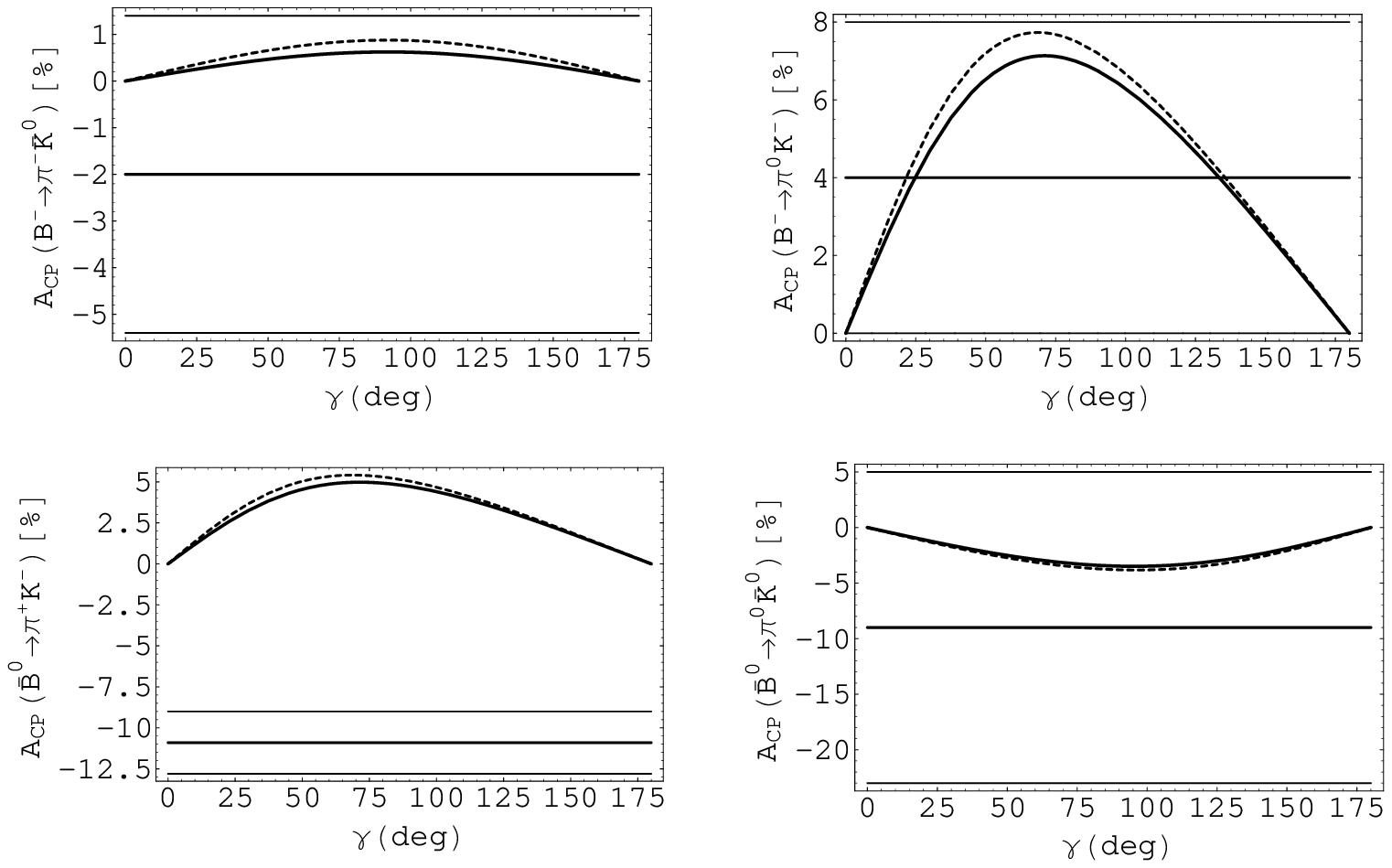}} \\
\scalebox{0.8}{\epsfig{file=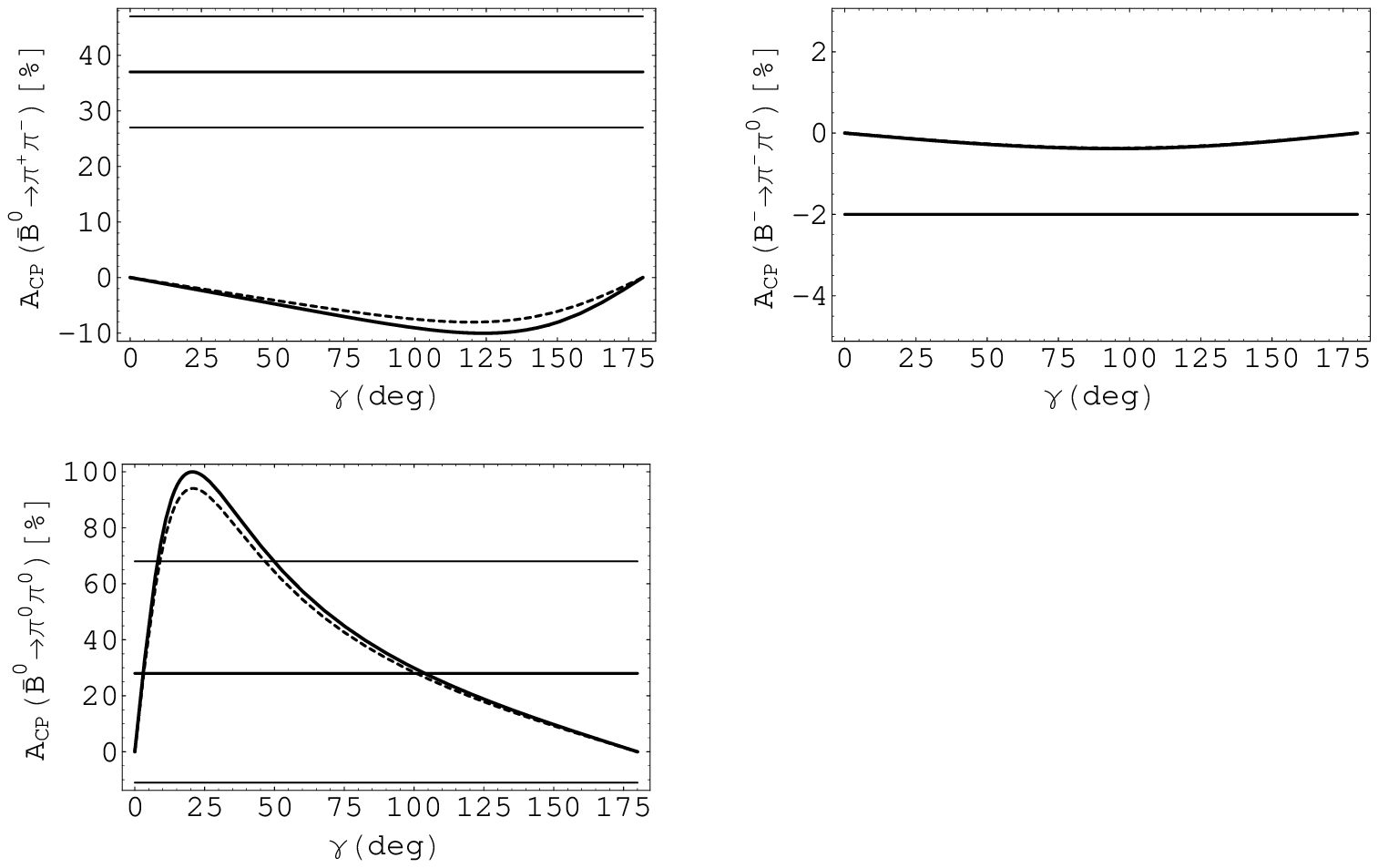}} \caption{ \label{fig:ACP1}
The $\gamma$ dependence of the $CP$ asymmetries without the
annihilation contributions. The meaning of the curves and the
horizontal solid lines is the same as in Fig.~\ref{fig:qcdf}.}
\end{center}
\end{figure}

\begin{figure}[t]
\begin{center}
\scalebox{0.8}{\epsfig{file=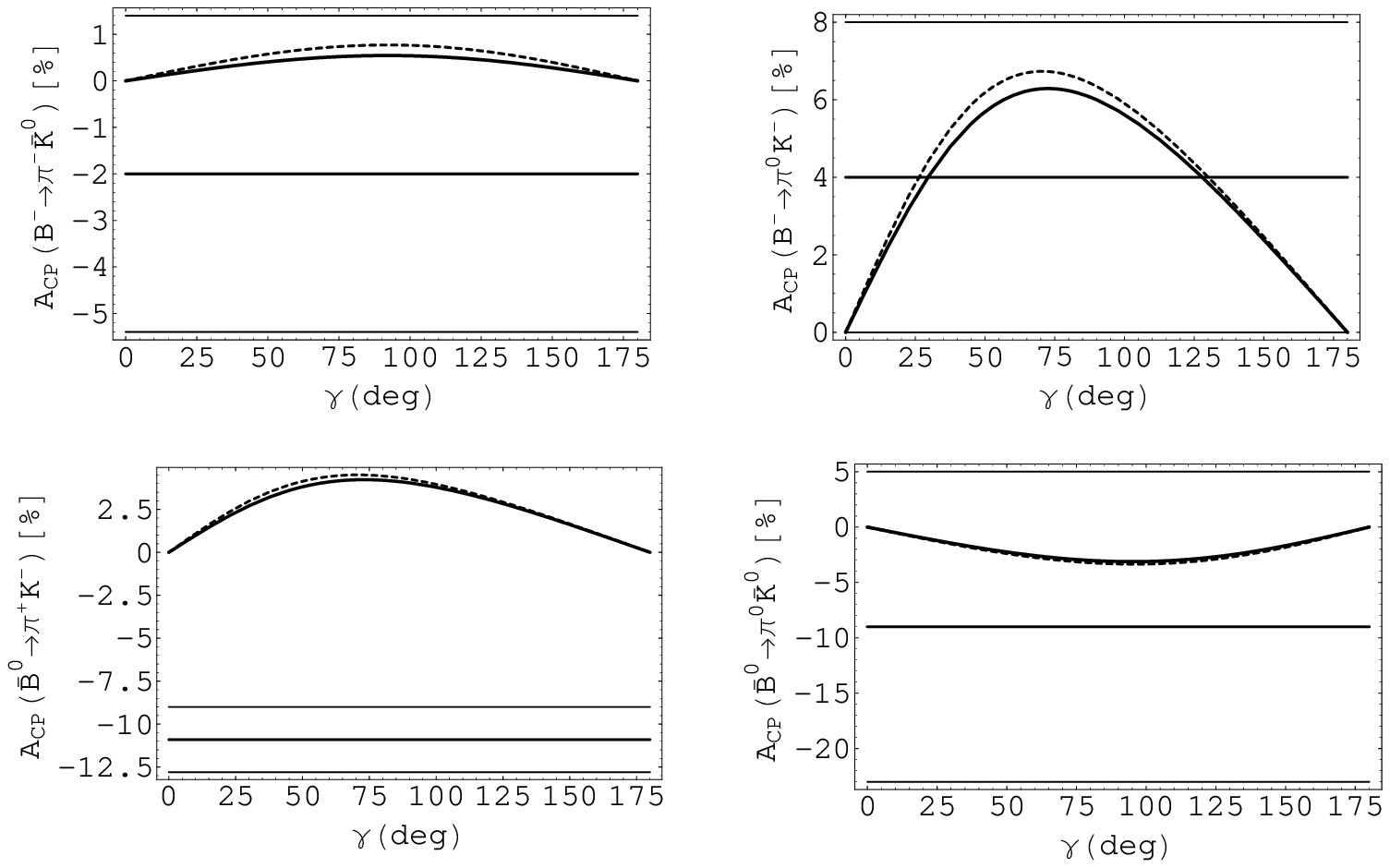}} \\
\scalebox{0.8}{\epsfig{file=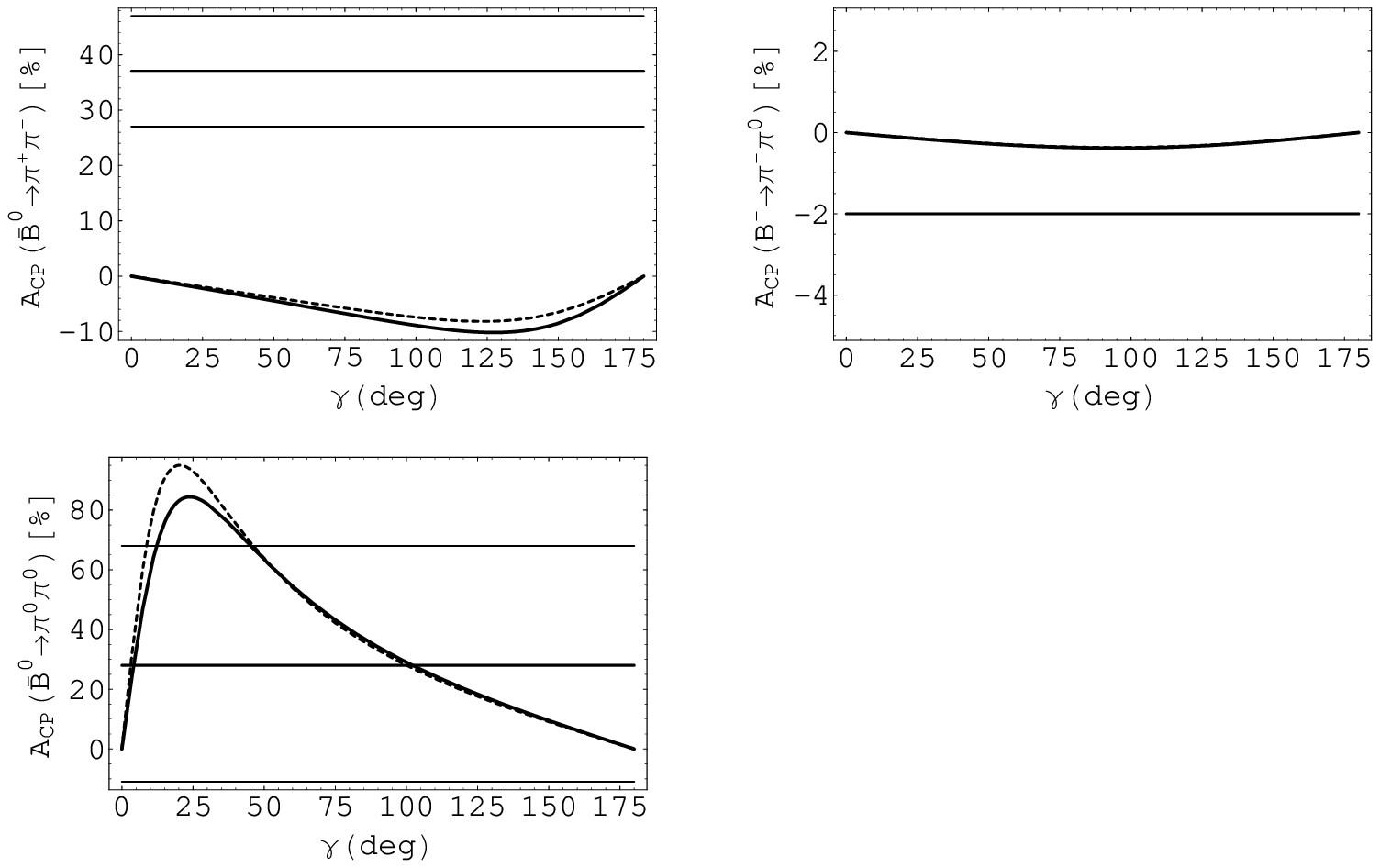}} \caption{\label{fig:ACP2}
The $\gamma$ dependence of the $CP$ asymmetries with the
annihilation contributions included. The meaning of the curves and
the horizontal solid lines is the same as in Fig.~\ref{fig:qcdf}.}
\end{center}
\end{figure}

From these two figures and the numerical results given in
Table~\ref{ACP}, we can see that:
\begin{itemize}

\item The direct $CP$ asymmetries for $B\to \pi K, \pi\pi$ decays
are predicted to be typically small with the QCDF formalism. This
could be well understood, since the direct $CP$ asymmetries are
proportional to the sines of the strong interaction phases, which
are usually suppressed by $\alpha_s$ and/or $\Lambda_{QCD}/m_b$
within the QCDF formalism. Due to a potentially large relative
phase between the QCD penguins and the coefficient $a_2$, the
$\overline{B}^0 \to \pi^0 \pi^0$ mode, however, is an exception to
this general rule. The direct $CP$ asymmetries for this mode is
predicted to be about $55\%$.

\item Although the individual Feynman diagram in
Fig.~\ref{penguinfig} carries large strong phase, the combining
contributions of these $b\to D g^\ast g^\ast$ strong penguin
diagrams contain only a relatively small one. Thus, these higher
order strong penguin contributions to the direct $CP$ asymmetries
are also small.

\item The theoretical predictions for $A_{CP}(\overline{B}^0 \to
\pi^{+} \pi^{-}) $ and $A_{CP}(\overline{B}^0 \to \pi^{+} K^{-})$
are quite smaller than the experimental data, particularly with
the opposite sign. How to accommodate these discrepancies in the
SM is still a challenge.

\end{itemize}

\section{Conclusions}

In this paper, we have revisited  the $B\to \pi K,\pi\pi$ decays
in the framework of QCDF with the $b\to D g^\ast g^\ast$ strong
penguin contributions included. The main conclusions of this paper
are:
\begin{enumerate}

\item For penguin-dominated $B\to \pi K$ decays, the higher order
strong penguin contributions induced by $b\to s g^\ast g^\ast$
transitions to the branching ratios are rather large. With our
input parameters, we find that these higher order strong penguin
contributions can give $\sim 30\%$ enhancement to the
corresponding branching ratios, and such an enhancement can
improve the consistency between the theoretical predictions and
the experimental data significantly .

\item For tree-dominated $B\to \pi\pi$ decays, the higher order
$b\to d g^\ast g^\ast$ contributions to the corresponding
branching ratios are quite small.

\item Because of large cancellations among  the $b\to D g^\ast
g^\ast$ strong penguin contributions, only a relatively small
strong phase is remained, So that the contributions have small
effects on predictions of the direct $CP$ asymmetries.

\item Since corrections of these higher order strong penguin
diagrams to the decay amplitudes are similar in nature, and hence
cancelled in the ratios between the corresponding branching
fractions, the patterns of the quantities $R$, $R_c$, $R_n$,
$R_{+-}$, and $R_{00}$ remain unaffected compared to the
next-to-leading order results. So we haven't found solution to the
``$\pi K$" puzzle. Our results indicate  that to resolve the
puzzle we may have to resort to new physics contributions through
the electroweak penguin sector as observed by Buras {\it et
al}.~\cite{Buras:2004th}.

\item The theoretical predictions for the branching ratios and the
direct $CP$ asymmetries still have large theoretical
uncertainties. The dominant errors are induced by the
uncertainties of the $F_0^{B \to \pi,K}(q^2)$ form factors,
strange quark mass $\overline{m}_s(\mu)$, and the CKM angle
$\gamma$.
\end{enumerate}

Although the results presented here have still large theoretical
uncertainties, the $b\to D g^\ast g^\ast$ strong penguin
contributions to two-body hadronic $B$-meson decays, particularly
to penguin-dominated modes, have been shown to be very important.
Further systematic studies on these higher order contributions to
charmless $B$ decays are therefore interesting and deserving.

\textit{Note added: After this work is finished,  we note an
interesting study of $\alpha_s^2$ corrections to $B\to \pi K,
\pi\pi$ decays has been carried out by Li, Mishima and
Sanda~\cite{lihn2} in PQCD formalism. However, the contributions
studied here as depicted by the Feynman diagrams in
Figs.~\ref{Q8gfig} and~\ref{penguinfig} are not included in their
paper.}

\section*{Acknowledgments}
The work is supported  by National Science Foundation under
contract No.10305003, Henan Provincial Foundation for Prominent
Young Scientists under contract No.0312001700 and
the NCET Program  sponsored by Ministry of Education, China.\\

\begin{appendix}
\begin{center}
\end{center}

\section*{Appendix A: Correction functions at next-to-leading order in $\alpha_s$}

In this appendix, we present the explicit form for the correction
functions appearing in the parameters $a_i$ and $b_i$. Details
about the calculation can be found in Refs.~\cite{bbns2,bbns3}.

\noindent -{\bf One-loop vertex corrections.} The vertex
parameters $V_i(M_2)$ result from the first four diagrams in
Fig.~\ref{asfig}, given by (with $M_2=\pi$, or $K$)
\begin{equation}\label{vertex}
   V_i(M_2) = \left\{\,\,
   \begin{array}{ll}
    {\displaystyle \int_0^1\!du\,\Phi_{M_2}(u)\,
     \Big[ 12\ln\frac{m_b}{\mu} - 18 + g(u) \Big]} \, & \qquad
     i=\mbox{1--4},9,10, \\[0.4cm]
   {\displaystyle \int_0^1\!du\,\Phi_{M_2}(u)\,
     \Big[ - 12\ln\frac{m_b}{\mu} + 6 - g(1-u) \Big]} \, & \qquad
     i=5,7, \\[0.4cm]
   {\displaystyle \int_0^1\!du\,\Phi_p^{M_2}(u)\,[-6\,]}
    \, & \qquad i=6,8,
   \end{array}\right.
\end{equation}
with
\begin{equation}\label{vertexfunction}
   g(u) = 3\left[ \frac{1-2u}{1-u}\ln u-i\pi \right].
\end{equation}
The scheme-dependent constants $-18$, $6$, $-6$ are specific to
the NDR scheme for $\gamma_5$. $\Phi_{M_2}$ and $\Phi_p^{M_2}$
denote the leading-twist and twist-3 LCDAs of the emitted meson
$M_2$, respectively.

\noindent -{\bf Penguin contractions.} The QCD and electro-weak
penguin parameters $P_{4,6}^p$ and $P_{8,10}^p$ arise from the
diagrams in Figs.~\ref{asfig}(e) and~\ref{asfig}(f) . Considering
the fact that there exist two distinct penguin contractions as
shown in Fig.~\ref{twocontraction}, these penguin contributions
can be written as
\begin{eqnarray}\label{penguin}
   P_4^p(M_2) &=& \frac{C_F\alpha_s}{4\pi N_c}\,\Bigg\{
    C_1 \!\left[ \frac43\ln\frac{m_b}{\mu}
    + \frac23 - G_{M_2}(s_p) \right]\! \Bigg.\nonumber\\
   &&\mbox{}+ \Bigg. C_3 \!\left[ \frac83\ln\frac{m_b}{\mu} + \frac43
    - G_{M_2}(0) - G_{M_2}(1) \right]\! \Bigg.\nonumber\\
   &&\mbox{}+ \Bigg. (C_4+C_6)\!
    \left[ \frac{4n_f}{3}\ln\frac{m_b}{\mu}
    - (n_f-2)\,G_{M_2}(0) - G_{M_2}(s_c) - G_{M_2}(1) \right]\! \Bigg.\nonumber\\
   &&\mbox{}- \Bigg. 2 C_{8g}^{\rm eff} \int_0^1 \frac{du}{1-u}\,
    \Phi_{M_2}(u) \Bigg\} \,,\\
    P_6^p(M_2) &=& \frac{C_F\alpha_s}{4\pi N_c}\,\Bigg\{
    C_1 \!\left[ \frac43\ln\frac{m_b}{\mu}
    + \frac23 - \hat G_{M_2}(s_p) \right]\! \Bigg.\nonumber\\
   &&\mbox{}+ \Bigg. C_3 \!\left[ \frac83\ln\frac{m_b}{\mu} + \frac43
    - \hat G_{M_2}(0) - \hat G_{M_2}(1) \right]\! \Bigg.\nonumber\\
   &&\mbox{}+ \Bigg. (C_4+C_6)\!
    \left[ \frac{4n_f}{3}\ln\frac{m_b}{\mu}
    - (n_f-2)\,\hat G_{M_2}(0) - \hat G_{M_2}(s_c) - \hat G_{M_2}(1)
    \right]\! \Bigg.\nonumber\\
   &&\mbox{}- \Bigg. 2 C_{8g}^{\rm eff} \Bigg\}\,,\\
   P_{10}^p(M_2) &=& \frac{\alpha}{9\pi N_c}\,\left\{
   (C_1+N_c C_2) \left[ \frac{4}{3}\ln\frac{m_b}{\mu} + \frac23
   - G_{M_2}(s_p) \right]\! \right.\nonumber\\
   &&\mbox{}- \left. 3 C_{7\gamma}^{\rm eff} \int_0^1 \frac{du}{1-u}\,\Phi_{M_2}(u)
   \right\} \,,\\
   P_8^p(M_2) &=& \frac{\alpha}{9\pi N_c}\,\left\{
   (C_1+N_c C_2) \left[ \frac{4}{3}\ln\frac{m_b}{\mu} + \frac23
   - \hat G_{M_2}(s_p) \right]\! - 3 C_{7\gamma}^{\rm eff} \right\}\,,
\end{eqnarray}
where $C_{F}=\frac{N_{c}^{2}-1}{2N_{c}}$, and $N_{c}=3$ is the
number of colors. $n_f=5$ is the number of light quark flavors.
The pole quark mass ratios, $s_u=0$, $s_c=(m_c/m_b)^2$, are
involved in the evaluation of these penguin diagrams. The function
$G_{M_2}(s)$ and $\hat G_{M_2}(s)$ are defined, respectively, by
\begin{eqnarray}
   G_{M_2}(s) &=& \int_0^1\!du\,G(s,1-u)\,\Phi_{M_2}(u)
   \,,\nonumber\\
   \hat G_{M_2}(s) &=& \int_0^1\!du\,G(s,1-u)\,\Phi_p^{M_2}(u) \,,
\end{eqnarray}
with
\begin{equation}\label{Gfunction}
G(s,u) = -4\int_0^1\!dx\,x\,(1-x) \ln[s-x(1-x)u-i\delta\,],
\end{equation}
where the term $i\delta$ is the ``$\epsilon$-prescription". The
interpretation of $\Phi_{M_2}$ and $\Phi_p^{M_2}$ is the same as
in the discussion of vertex corrections.

\begin{figure}[t]
\epsfxsize=10cm \centerline{\epsffile{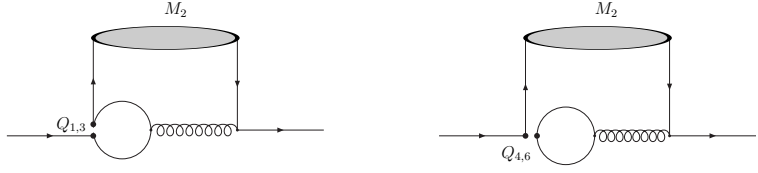}}
\centerline{\parbox{14cm}{\caption{\label{twocontraction} Two
different penguin contractions.}}}
\end{figure}

\noindent -{\bf Hard spectator interactions.} The parameters
$H_i(M_1 M_2)$ originate from the hard gluon exchange between the
meson $M_2$ and the spectator quark~(corresponding to the last two
diagrams in Fig.~\ref{asfig}) with the results given by
\begin{eqnarray}\label{hardspectator1}
   H_i(M_1M_2)
   &=& \frac{f_B \,f_{M_1}}{(m_B^2-m_{M_1}^2)\,F_0^{B\to M_1}(m_{M_2}^2)}\,
    \int_0^1 \frac{d\xi}{\xi}\,\Phi_1^B(\xi)\,\nonumber\\
   &&\mbox{} \times \int_0^1 \!du \int_0^1 \!dv \,\left[
    \frac{\Phi_{M_2}(u) \,\Phi_{M_1}(v)}{(1-u)\,(1-v)}\,
   + r_\chi^{M_1}\,\frac{\Phi_{M_2}(u)\Phi_p^{M_1}(v)}{u\,(1-v)}
   \right]\,,
\end{eqnarray}
for $i=1$--4,9,10,
\begin{eqnarray}\label{hardspector2}
   H_i(M_1M_2)
   &=& -\frac{f_B \,f_{M_1}}{(m_B^2-m_{M_1}^2)\,F_0^{B\to M_1}(m_{M_2}^2)}\,
    \int_0^1 \frac{d\xi}{\xi}\,\Phi_1^B(\xi)\,\nonumber\\
   &&\mbox{} \times \int_0^1 \!du \int_0^1 \!dv \,\left[
    \frac{\Phi_{M_2}(u) \,\Phi_{M_1}(v)}{u\,(1-v)}\,
   + r_\chi^{M_1}\,\frac{\Phi_{M_2}(u)\Phi_p^{M_1}(v)}{(1-u)\,(1-v)}
   \right]\,,
\end{eqnarray}
for $i=5,7$, and $H_i(M_1M_2)=0$ for $i=6,8$. In these results
$\Phi_1^B(\xi)$ is the leading twist LCDAs of the $B$ meson as
defined by Eq.~(\ref{Bprojector}).

\noindent -{\bf Weak annihilation contributions.} The basic
building blocks for annihilation contributions originate from
Fig.~\ref{annhfig} and given by (omitting the argument $M_1 M_2$
for brevity)
\begin{eqnarray}\label{blocks}
   A_1^i &=& \pi\alpha_s \int_0^1\! du dv\,
    \left\{ \Phi_{M_2}(u)\,\Phi_{M_1}(v)
    \left[ \frac{1}{v(1-u\bar v)} + \frac{1}{\bar u^2 v} \right]
    + r_\chi^{M_1} r_\chi^{M_2}\,\Phi_p^{M_2}(u)\,\Phi_p^{M_1}(v)\,
     \frac{2}{\bar u v} \right\} ,\nonumber\\
   A_1^f &=& 0 \,, \nonumber\\
   A_2^i &=& \pi\alpha_s \int_0^1\! du dv\,
    \left\{ \Phi_{M_2}(u)\,\Phi_{M_1}(v)
    \left[ \frac{1}{\bar u(1-u\bar v)} + \frac{1}{\bar u v^2} \right]
    + r_\chi^{M_1} r_\chi^{M_2}\,\Phi_p^{M_2}(u)\,\Phi_p^{M_1}(v)\,
     \frac{2}{\bar u v} \right\} ,\nonumber\\
   A_2^f &=& 0 \,, \nonumber\\
   A_3^i &=& \pi\alpha_s \int_0^1\! du dv\,
    \left\{r_\chi^{M_1}\,\Phi_{M_2}(u)\,\Phi_p^{M_1}(v)\,
    \frac{2\bar v}{\bar u v(1-u\bar v)}
    - r_\chi^{M_2}\,\Phi_{M_1}(v)\,\Phi_p^{M_2}(u)\,
    \frac{2u}{\bar u v(1-u\bar v)} \right\} , \nonumber\\
   A_3^f &=& \pi\alpha_s \int_0^1\! du dv\,
    \left\{r_\chi^{M_1}\,\Phi_{M_2}(u)\,\Phi_p^{M_1}(v)\,
    \frac{2(1+\bar u)}{\bar u^2 v}
    +  r_\chi^{M_2}\,\Phi_{M_1}(v)\,\Phi_p^{M_2}(u)\,
    \frac{2(1+v)}{\bar u v^2} \right\}\,,
\end{eqnarray}
where the superscripts `$i$' and `$f$' refer to gluon emission
from the initial and final-state quarks, respectively. The
subscript `$k$' refers to one of the three possible Dirac
structures $\Gamma_1\otimes\Gamma_2$, i.e., $k=1$ for
$(V-A)\otimes(V-A)$, $k=2$ for $(V-A)\otimes(V+A)$, and $k=3$ for
$(-2)(S-P)\otimes(S+P)$.

Considering the off-shellness of the gluon in Fig.~\ref{asfig} and
Fig.~\ref{annhfig}, it is reasonable to evaluate the vertex and
penguin corrections at the scale $\mu\sim m_b$, while the hard
spectator scattering and the weak annihilations contributions at
the scale $\mu_h=\sqrt{\Lambda_h\,\mu}$ with $\Lambda_h=0.5\,{\rm
GeV}$.

\section*{Appendix B: Analytic expressions for the $\Delta_i$ functions}

In the NDR scheme, after performing the loop-momentum integration,
we can present the analytic expressions for the $\Delta_i$
functions appearing in Eqs.~(\ref{TPfunction})
and~(\ref{TMfunction}) as
\begin{eqnarray}
\Delta i_{5} &=&
-\Gamma(\frac{\epsilon}{2})\,(4\,\pi\mu^2)^{\frac{\epsilon}{2}}\,
\int_0^1\!dx\int_0^{1-x}\!dy\,C^{-1-\frac{\epsilon}{2}}\,\left[
  2\,m_q^2\,\epsilon -
  2\,m_q^2\,x\,\epsilon - 2\,k^2\,x^2\,\epsilon +
  2\,k^2\,x^3\,\epsilon \right.\,\nonumber\\
&&\mbox\, \left. \qquad \qquad \qquad + 2\,p^2\,y\,\epsilon  -
  2\,p^2\,y^2\,\epsilon + 2\,p^2\,x\,y^2\,\epsilon +
  4\,(k \cdot p)\, x\,y\,\epsilon -
  4\,(k \cdot p) \, x^2\,y\,\epsilon \, \right.\,\nonumber\\
&&\mbox\, \left. \qquad \qquad \qquad + 4\,C - 12\,C\,x -
  4\,C\,x\,\epsilon \,\right]\,,\\
\Delta i_{6} &=&
\Gamma(\frac{\epsilon}{2})\,(4\,\pi\mu^2)^{\frac{\epsilon}{2}}\,
\int_0^1\!dx\int_0^{1-x}\!dy\,C^{-1-\frac{\epsilon}{2}}\,\left[
  2\,m_q^2\,\epsilon +
  2\,k^2\,x\,\epsilon  - 2\,k^2\,x^2\,\epsilon  -
  2\,m_q^2\,y\,\epsilon \, \right.\,\nonumber\\
&&\mbox\, \left. \qquad \qquad \qquad +
  2\,k^2\,x^2\,y\,\epsilon  - 2\,p^2\,y^2\,\epsilon  +
  2\,p^2\,y^3\,\epsilon  +
  4\,(k \cdot p)\,x\,y\,\epsilon -
  4\,(k \cdot p)\,x\,y^2\,\epsilon \, \right.\,\nonumber\\
&&\mbox\, \left. \qquad \qquad \qquad + 4\,C - 12\,C\,y -
  4\,C\,y\,\epsilon \,\right]\,,\\
\Delta i_{23} &=& 4\,
\Gamma(\frac{\epsilon}{2})\,(4\,\pi\mu^2)^{\frac{\epsilon}{2}}\,(k
\cdot p)\,\epsilon\,
\int_0^1\!dx\int_0^{1-x}\!dy\,C^{-1-\frac{\epsilon}{2}}\,x\,y\,,\\
\Delta i_{24} &=& 4\,
\Gamma(\frac{\epsilon}{2})\,(4\,\pi\mu^2)^{\frac{\epsilon}{2}}\,(k
\cdot p)\,\epsilon\,
\int_0^1\!dx\int_0^{1-x}\!dy\,C^{-1-\frac{\epsilon}{2}}\,y\,(1-y)\,,\\
\Delta i_{25} &=& -4\,
\Gamma(\frac{\epsilon}{2})\,(4\,\pi\mu^2)^{\frac{\epsilon}{2}}\,(k
\cdot p)\,\epsilon\,
\int_0^1\!dx\int_0^{1-x}\!dy\,C^{-1-\frac{\epsilon}{2}}\,x\,(1-x)\,,\\
\Delta i_{26} &=& -\Delta i_{23}\,,\\
\Delta i_{2} &=&
\Gamma(\frac{\epsilon}{2})\,(4\,\pi\mu^2)^{\frac{\epsilon}{2}}\,
\int_0^1\!dx\int_0^{1-x}\!dy\,C^{-1-\frac{\epsilon}{2}}\,\left[
  2\,m_q^2\,\epsilon  -
  2\,m_q^2\,x\,\epsilon  -
  2\,k^2\,x^2\,\epsilon  + 2\,k^2\,x^3\,\epsilon \, \right.\,\nonumber\\
&&\mbox\, \left. \qquad \qquad \qquad +
  2\,p^2\,y\,\epsilon  - 2\,p^2\,y^2\,\epsilon  +
  2\,p^2\,x\,y^2\,\epsilon  +
  4\,(k \cdot p)\,x\,y\,\epsilon -
  4\,(k \cdot p)\,x^2\,y\,\epsilon \, \right.\,\nonumber\\
&&\mbox\, \left. \qquad \qquad \qquad +
  4\,C - 4\,C\,x - 4\,C\,\epsilon  + 4\,C\,x\,\epsilon\,\right]\,,\\
\Delta i_{3} &=&
-\Gamma(\frac{\epsilon}{2})\,(4\,\pi\mu^2)^{\frac{\epsilon}{2}}\,
\int_0^1\!dx\int_0^{1-x}\!dy\,C^{-1-\frac{\epsilon}{2}}\,\left[
  2\,m_q^2\,\epsilon +
  2\,k^2\,x\,\epsilon  -
  2\,k^2\,x^2\,\epsilon  - 2\,m_q^2\,y\,\epsilon \, \right.\,\nonumber\\
&&\mbox\, \left. \qquad \qquad \qquad +
  2\,k^2\,x^2\,y\,\epsilon  - 2\,p^2\,y^2\,\epsilon  +
  2\,p^2\,y^3\,\epsilon  +
  4\,(k \cdot p)\,x\,y\,\epsilon -
  4\,(k \cdot p)\,x\,y^2\,\epsilon \, \right.\,\nonumber\\
&&\mbox\, \left. \qquad \qquad \qquad +
  4\,C - 4\,C\,y -4\,C\,\epsilon + 4\,C\,y\,\epsilon\,\right]\,,\\
\Delta i_{8} &=&
-\Gamma(\frac{\epsilon}{2})\,(4\,\pi\mu^2)^{\frac{\epsilon}{2}}\,
\int_0^1\!dx\int_0^{1-x}\!dy\,C^{-1-\frac{\epsilon}{2}}\,\left[
  2\,m_q^2\,\epsilon +
  2\,m_q^2\,x\,\epsilon  +
  2\,k^2\,x^2\,\epsilon  - 2\,k^2\,x^3\,\epsilon \, \right.\,\nonumber\\
&&\mbox\, \left. \qquad \qquad \qquad +
  2\,p^2\,y\,\epsilon  - 2\,p^2\,y^2\,\epsilon  -
  2\,p^2\,x\,y^2\,\epsilon  +
  4\,(k \cdot p)\,x^2\,y\,\epsilon \, \right.\,\nonumber\\
&&\mbox\, \left. \qquad \qquad \qquad +
  4\,C + 4\,C\,x - 4\,C\,\epsilon  - 4\,C\,x\,\epsilon\,\right]\,,\\
\Delta i_{9} &=&
-\Gamma(\frac{\epsilon}{2})\,(4\,\pi\mu^2)^{\frac{\epsilon}{2}}\,
\int_0^1\!dx\int_0^{1-x}\!dy\,C^{-1-\frac{\epsilon}{2}}\,\left[
  2\,m_q^2\,\epsilon +
  2\,k^2\,x\,\epsilon  - 2\,k^2\,x^2\,\epsilon  -
  2\,m_q^2\,y\,\epsilon  \, \right.\,\nonumber\\
&&\mbox\, \left. \qquad \qquad \qquad -
  4\,k^2\,x\,y\,\epsilon  +
  2\,k^2\,x^2\,y\,\epsilon  - 2\,p^2\,y^2\,\epsilon  +
  2\,p^2\,y^3\,\epsilon  \, \right.\,\nonumber\\
&&\mbox\, \left. \qquad \qquad \qquad -
  4\,(k \cdot p)\,y\,\epsilon +
  4\,(k \cdot p)\,x\,y\,\epsilon +
  4\,(k \cdot p)\,y^2\,\epsilon  -
  4\,(k \cdot p)\,x\,y^2\,\epsilon \, \right.\,\nonumber\\
&&\mbox\, \left. \qquad \qquad \qquad +
  4\,C - 4\,C\,y - 4\,C\,\epsilon  + 4\,C\,y\,\epsilon\,\right]\,,\\
\Delta i_{11} &=&
\Gamma(\frac{\epsilon}{2})\,(4\,\pi\mu^2)^{\frac{\epsilon}{2}}\,
\int_0^1\!dx\int_0^{1-x}\!dy\,C^{-1-\frac{\epsilon}{2}}\,\left[
  2\,m_q^2\,\epsilon -
  2\,m_q^2\,x\,\epsilon  -
  2\,k^2\,x^2\,\epsilon  + 2\,k^2\,x^3\,\epsilon \, \right.\,\nonumber\\
&&\mbox\, \left. \qquad \qquad \qquad +
  2\,p^2\,y\,\epsilon  - 4\,p^2\,x\,y\,\epsilon  -
  2\,p^2\,y^2\,\epsilon  + 2\,p^2\,x\,y^2\,\epsilon \, \right.\,\nonumber\\
&&\mbox\, \left. \qquad \qquad \qquad  -
  4\,(k \cdot p)\,x\,\epsilon +
  4\,(k \cdot p)\,x^2\,\epsilon +
  4\,(k \cdot p)\,x\,y\,\epsilon -
  4\,(k \cdot p)\,x^2\,y\,\epsilon \, \right.\,\nonumber\\
&&\mbox\, \left. \qquad \qquad \qquad +
  4\,C - 4\,C\,x - 4\,C\,\epsilon  + 4\,C\,x\,\epsilon\,\right]\,,\\
\Delta i_{12} &=&
\Gamma(\frac{\epsilon}{2})\,(4\,\pi\mu^2)^{\frac{\epsilon}{2}}\,
\int_0^1\!dx\int_0^{1-x}\!dy\,C^{-1-\frac{\epsilon}{2}}\,\left[
  2\,m_q^2\,\epsilon +
  2\,k^2\,x\,\epsilon  -
  2\,k^2\,x^2\,\epsilon  + 2\,m_q^2\,y\,\epsilon  \, \right.\,\nonumber\\
&&\mbox\, \left. \qquad \qquad \qquad-
  2\,k^2\,x^2\,y\,\epsilon  + 2\,p^2\,y^2\,\epsilon  -
  2\,p^2\,y^3\,\epsilon  +
  4\,(k \cdot p)\,x\,y^2\,\epsilon \, \right.\,\nonumber\\
&&\mbox\, \left. \qquad \qquad \qquad +
  4\,C + 4\,C\,y - 4\,C\,\epsilon  - 4\,C\,y\,\epsilon\,\right]\,,\\
\Delta i_{15} &=&
8\,\Gamma(\frac{\epsilon}{2})\,(4\,\pi\mu^2)^{\frac{\epsilon}{2}}\,(k
\cdot p)\,\epsilon\,
\int_0^1\!dx\int_0^{1-x}\!dy\,C^{-1-\frac{\epsilon}{2}}\,(1-x)\,x^2\,,\\
\Delta i_{16} &=&
4\,\Gamma(\frac{\epsilon}{2})\,(4\,\pi\mu^2)^{\frac{\epsilon}{2}}\,(k
\cdot p)\,\epsilon\,
\int_0^1\!dx\int_0^{1-x}\!dy\,C^{-1-\frac{\epsilon}{2}}\,\,x\,(1-x)\,(1-2\,y)\,,\\
\Delta i_{17} &=&
-4\,\Gamma(\frac{\epsilon}{2})\,(4\,\pi\mu^2)^{\frac{\epsilon}{2}}\,(k
\cdot p)\,\epsilon\,
\int_0^1\!dx\int_0^{1-x}\!dy\,C^{-1-\frac{\epsilon}{2}}\,x\,y\,(1-2\,x)\,,\\
\Delta i_{18} &=&
-4\,\Gamma(\frac{\epsilon}{2})\,(4\,\pi\mu^2)^{\frac{\epsilon}{2}}\,(k
\cdot p)\,\epsilon\,
\int_0^1\!dx\int_0^{1-x}\!dy\,C^{-1-\frac{\epsilon}{2}}\,y\,(1-x-y+2\,x\,y)\,,\\
\Delta i_{19} &=&
4\,\Gamma(\frac{\epsilon}{2})\,(4\,\pi\mu^2)^{\frac{\epsilon}{2}}\,(k
\cdot p)\,\epsilon\,
\int_0^1\!dx\int_0^{1-x}\!dy\,C^{-1-\frac{\epsilon}{2}}\,x\,(1-x-y+2\,x\,y)\,,\\
\Delta i_{20} &=&
-4\,\Gamma(\frac{\epsilon}{2})\,(4\,\pi\mu^2)^{\frac{\epsilon}{2}}\,(k
\cdot p)\,\epsilon\,
\int_0^1\!dx\int_0^{1-x}\!dy\,C^{-1-\frac{\epsilon}{2}}\,(1-2\,x)\,y\,(1-y)\,,\\
\Delta i_{21} &=&
4\,\Gamma(\frac{\epsilon}{2})\,(4\,\pi\mu^2)^{\frac{\epsilon}{2}}\,(k
\cdot p)\,\epsilon\,
\int_0^1\!dx\int_0^{1-x}\!dy\,C^{-1-\frac{\epsilon}{2}}\,x\,y\,(1-2\,y)\,,\\
\Delta i_{22} &=&
-8\,\Gamma(\frac{\epsilon}{2})\,(4\,\pi\mu^2)^{\frac{\epsilon}{2}}\,(k
\cdot p)\,\epsilon\,
\int_0^1\!dx\int_0^{1-x}\!dy\,C^{-1-\frac{\epsilon}{2}}\,(1-y)\,y^2\,,
\end{eqnarray}
where the parameter $C$ is defined by
\begin{equation}
C = m_q^2 - x\,(1-x)\,k^2 -y\,(1-y)\,p^2 -2\,x\,y \,(k \cdot p)
-i\,\delta.
\end{equation}
with $m_q$ being the quark mass in the Fermion loops.

For $B$ meson decaying into two light energetic hadronic final
states, the characteristic scale for the quark momentum of the
final-state meson constituents is of order $m_b$, whereas the
momentum of the spectator quark from the $B$ meson is of order
$\Lambda_{\rm QCD}$. Assuming that the off-shell gluon with index
$(\nu,b,p)$ is connected with the spectator quark in the $B$
meson, at leading power in $\Lambda_{\rm QCD}/m_b$, the $\Delta i$
functions given above can then be simplified greatly. After
subtracting the regulator $\epsilon$ using the $\overline{\rm MS}$
scheme and performing the Feynman parameter integrals, we
get~(here we give only the relevant $\Delta i$ functions needed in
this paper; details for the others can be found in
Ref.~\cite{yang:phiXs})
\begin{eqnarray}
\Delta i_{5} &=& 2 + \frac{2\,r_1}{r_3}\,\left[G_0(r_1) - G_0(r_1+
r_3)\,\right]\, - \frac{4}{r_3}\,\left[G_{-1}(r_1) -
G_{-1}(r_1 + r_3)\,\right]\,,\\
\Delta i_{6} &=& -2 - \frac{4}{r_3} +
  \frac{2\,r_1\left( 1 +r_3 \right) }{r_3^2}\,G_0(r_1) -
  \frac{2\,\left( r_1 + r_3 + r_1\,r_3 \right) }{r_3^2}\,G_0(r_1 + r_3)\,\nonumber\\
&&\mbox{} +
  \frac{4}{r_3}\left[G_{-1}(r_1) - G_{-1}(r_1+r_3)\,\right]\, -
  \frac{\left( 4 - r_1 \right) \,r_1}{r_3^2}\,T_0(r_1) \,\nonumber\\
&&\mbox{} +
  \frac{\left( 4 - r_1 - r_3 \right) \,\left(  r_1 + r_3 \right)
  }{r_3^2}\,T_0( r_1 + r_3)\,,\\
\Delta i_{23} &=& -2 - \frac{2\,r_1}{r_3}\,\left[G_0(r_1) -
G_0(r_1+ r_3)\,\right]\, + \frac{4}{r_3}\,\left[G_{-1}(r_1) -
G_{-1}(r_1+ r_3)\,\right]\,,\\
\Delta i_{26} &=& -\Delta i_{23}\,,\\
\Delta i_{2} &=&-\frac{22}{9}+ \frac{8}{3}\,\ln \frac{\mu}{m_c} -
  \frac{2\,\left(8+ r_1 \right) }{3\,r_3}\,G_0(r_1) +
  \frac{2\,\left( 8+ r_1 - 2\,r_3 \right) }{3\,r_3}\,G_0(r_1 + r_3)\,\nonumber\\
&&\mbox +
  \frac{4}{r_3}\left[G_{-1}(r_1) - G_{-1}(r_1+r_3)\,\right]\,,\\
\Delta i_{3} &=& \frac{22}{9} + \frac{12}{r_3} +
\frac{4\,r_1}{3\,r_3} - \frac{8}{3}\,\ln \frac{\mu}{m_c} -
  \frac{2\,\left( 7\,r_1 - r_3 - 3\,r_1\,r_3 + 2\,r_1^2 - 2\,r_3^2 \right) }
  {3\,r_3^2}\,G_0(r_1 + r_3)\,\nonumber\\
&& \mbox +
  \frac{2\,r_1\left( 7 + 2\,r_1 -
  3\,r_3 \right) }{3\,r_3^2}\,G_0(r_1) -
  \frac{4\,\left( 2\,r_1 + r_3 \right) }{r_3^2}\,
  \left[G_{-1}(r_1) - G_{-1}(r_1+r_3)\,\right]\,\nonumber\\
&&\mbox +
  \frac{3\,\left( 4 - r_1 \right) \,r_1}{r_3^2}\,T_0(r_1)-
  \frac{3\,\left( 4 - r_1 - r_3 \right) \,\left(  r_1 + r_3 \right)
  }{r_3^2}\,T_0( r_1 + r_3)\,,\\
\Delta i_{8} &=& \frac{32}{9} - \frac{16}{3}\,\ln \frac{\mu}{m_c}-
  \frac{8\left(2 + r_1 \right) }{3\,r_3}\,G_0(r_1)+
  \frac{8\,\left(2 + r_1 +r_3 \right) }{3\,r_3}\,G_0(r_1 + r_3)\,,\\
\Delta i_{12} &=& -\frac{32}{9} + \frac{12}{r_3} +
  \frac{4\,r_1}{3\,r_3} + \frac{16}{3}\,\ln \frac{\mu}{m_c}+
  \frac{2\,r_1\left( 7 + 2\,r_1 + 6\,r_3 \right) }{3\,r_3^2}\,
  G_0(r_1) \,\nonumber\\
&& -
  \frac{2\,\left( 2\,r_1^2 - r_3\,\left( 1 - 4\,r_3 \right)  +
   r_1\,\left( 7 + 6\,r_3 \right)  \right) }{3\,r_3^2} \,G_0(r_1 + r_3)\,\nonumber\\
&& -
  \frac{8\,r_1}{r_3^2}\,\left[G_{-1}(r_1) - G_{-1}(r_1+r_3)\,\right] +
  \frac{3\,\left( 4 - r_1 \right) \,r_1}{r_3^2}\,T_0(r_1) \,\nonumber\\
&& -
  \frac{3\,\left( 4 - r_1 - r_3 \right) \,\left( r_1+ r_3 \right) \,
  }{r_3^2}\,T_0(r_1+r_3) \,\\
\Delta i_{17} &=& \frac{2}{3} + \frac{2\,\left(8+r_1
\right)}{3\,r_3}\,G_{0}(r_1)
 -\frac{2}{3}\,\left(\frac{8+r_1}{r_3}+\frac{4}{r_1+r_3}\right)\,
 G_{0}(r_1 +r_3 ) \nonumber\\
 && -
 \frac{4}{r_3}\,\left[G_{-1}(r_1 )-G_{-1}(r_1 +r_3 ) \,\right]\,,\\
\Delta i_{21} &=& -\frac{2}{3} -\frac{16}{r_3}-\frac{8\,r_1
}{3\,r_3} +\frac{2\,r_1\left( 4\,r_1^2 + 3\,r_3\left( 8 + r_3
\right)  + r_1\left( 20 + 7\,r_3 \right)  \right) }
  {3\,r_3^2\,\left( r_1 + r_3\right) }\,G_{0}(r_1 + r_3 )\,\nonumber \\
&& -
  \frac{2\,r_1\,\left( 20 + 4\,r_1 + 3\,r_3 \right)}{3\,r_3^2}\,G_{0}(r_1 )+
  \frac{4\left( 4\,r_1 + r_3 \right) }{r_3^2}\,\left[ G_{-1}(r_1)- G_{-1}(r_1+r_3 )\right]\,\nonumber\\
&& -
  \frac{4\left( 4 - r_1 \right) \,r_1}{r_3^2}\,T_0(r_1) +
  \frac{4\left( 4 - r_1 - r_3 \right) \,\left( r_1+ r_3 \right) \,
  }{r_3^2}\,T_0(r_1+r_3) \,.
\end{eqnarray}
where we have introduced the notations $r_{1}=k^2/m_q^2$,
$r_{2}=p^2/m_q^2$ and $r_{3}=2\,(k\cdot p)/m_q^2$, with $m_q=m_c$
or $m_b$. For light $u,d,s$ quark loops, these $\Delta i$
functions can be evaluated straightforwardly.

The functions $G_i (t)~(i=-1,0)$ are defined by
\begin{equation}
G_{i}(t)=\int^1_0 \!dx\, x^i
\ln\left[1-x\,(1-x)\,t-i\delta\,\right]\,.
\end{equation}
The explicit form for $G_{-1,0}(t)$ could be found in
Ref.~\cite{greub}.

In addition, we have also introduced the function $T_i(t)$, which
is defined by
\begin{equation}
T_{i}(t)=\int^1_0 \!dx \frac{x^i}{1-x\,(1-x)\,t-i\delta}\,.
\end{equation}
The explicit form for $T_0(t)$ is given by~\cite{yang:phiXs}
\begin{eqnarray}
T_{0}(t) &=& \left\{
\begin{array}{cc}
\frac{ 4\,\arctan\sqrt{\frac{t}{4-t}}
}{\sqrt{t\,(4-t)}} ; &0\leq t \leq 4 \\
\frac{2i\,\pi+ 2\ln ( \sqrt{t}-\sqrt{t-4} )-2\ln (
\sqrt{t}+\sqrt{t-4})} {\sqrt{t\,(t-4)}};  &t > 4.
\end{array} \right.
\end{eqnarray}

\end{appendix}

\end{document}